\documentclass[preprint, number]{elsarticle}

\usepackage{hyperref}
\usepackage{subfigure}
\usepackage{amssymb,amsthm}
\usepackage{amsmath}
\usepackage{lscape}
\usepackage{threeparttable}
\usepackage{multirow}
\usepackage{color}
\usepackage{algorithm}
\usepackage{algorithmic}

\biboptions{compress}

\usepackage{setspace}
\usepackage{array}

\begin{document}
\begin{frontmatter}

\title{Price graphs: Utilizing the structural information of financial time series for stock prediction}

\author[mymainaddress]{Junran Wu}
\author[mymainaddress]{Ke Xu}
\author[mymainaddress]{Xueyuan Chen}
\author[mysecondaddress]{Shangzhe Li}

\author[mythirdaddress]{Jichang Zhao\corref{mycorrespondingauthor}}
\cortext[mycorrespondingauthor]{Corresponding author:}
\ead{jichang@buaa.edu.cn}

\address[mymainaddress]{State Key Lab of Software Development Environment, Beihang University}
\address[mysecondaddress]{School of Mathematics Science, Beihang University}
\address[mythirdaddress]{School of Economics and Management, Beihang University}

\begin{abstract}
Great research efforts have been devoted to exploiting deep neural networks in stock prediction. While long-term dependencies and chaotic property are still two major issues that lower the performance of state-of-the-art deep learning models in forecasting future price trends. In this study, we propose a novel framework to address both issues. Specifically, in terms of transforming time series into complex networks, we convert market price series into graphs. Then, structural information, referring to associations among temporal points and the node weights, is extracted from the mapped graphs to resolve the problems regarding long-term dependencies and the chaotic property. We take graph embeddings to represent the associations among temporal points as the prediction model inputs. Node weights are used as a priori knowledge to enhance the learning of temporal attention. The effectiveness of our proposed framework is validated using real-world stock data, and our approach obtains the best performance among several state-of-the-art benchmarks. Moreover, in the conducted trading simulations, our framework further obtains the highest cumulative profits. Our results supplement the existing applications of complex network methods in the financial realm and provide insightful implications for investment applications regarding decision support in financial markets.
\end{abstract}

\begin{keyword}
\texttt stock prediction \sep complex network \sep time series graph \sep graph embedding \sep structure information
\end{keyword}

\end{frontmatter}


\section{Introduction}
\label{sec:intro}
Financial time series prediction, in particular the stock prediction, which aims at forecasting the future trends of stock prices, is one of the key foundational techniques in investment and has attracted tremendous attention from various fields \citep{ou1989financial, qin2017dual, wu2020predicting}.
Various categories of methods (e.g., technical analysis, machine learning and deep learning models) and data sources (e.g., historical market price data, common funds and financial news) have been adopted for stock prediction \citep{blume1994market, qin2017dual, wu2020predicting, wu2020adaptive}, via the modeling of the relationship between the historical behavior and future trend \citep{fawaz2019deep}. 
While most of these traditional efforts, though equipped with state-of-the-art prediction algorithms, only concentrate on the structures among time series, the structures within time series are hardly investigated for stock prediction. 
With increasing literature leveraging complex network methods for characterizing dynamical systems derived from time series, complex networks have already served as promising and versatile tools for adopting structural information embedded within the temporal points of time series and providing new insights for stock prediction \citep{wang2012visibility, zou2019complex}.

In the current literature regarding stock prediction, without glimpsing at the structural information of stock time series, most research efforts have been devoted to various types of Internet information sources and dynamic indicators derived from stock prices \citep{blume1994market, wu2020predicting}. Currently, deep learning has become an effective method to refine multiaspect features from complex financial time series. Plenty of deep learning frameworks have been proposed in the literature with the aim of predicting asset prices \citep{qin2017dual, fawaz2019deep}.
To capture the interactions among multiple variables, attention mechanisms have also been deployed for time series forecasting \citep{qin2017dual, hao2020new, chen2021novel}.
However, the long-term dependencies of financial time series in experiments still have not been fully captured due to the complex temporal evolution of the interactions among all the temporal points. The value of a data point at time $t$ is not likely independent of its temporally neighboring points or historical points which are far before $t$. Such associations are denoted as short-term and long-term dependencies, respectively in deep learning \cite{bengio1994learning, hao2020new}.
In financial realm, it is analogously vital to utilize the dependencies among the values in the given series, and the long-term memories have also been uncovered and validated on daily, weekly, monthly, and annual stock returns \cite{lo1991long}. On the other hand, financial time series often represent chaotic and complex price movement behaviors \citep{farmer1987predicting}. As a result, stock prices are shown in non-stationary time series, and have abrupt changes or unexpected reversals which are taken as outliers in modeling and undermine the generalization ability of learning models. In particular, the chaotic property seriously challenges these models and makes them assign incorrect weights to points that are unhelpful for forecasting further trends. Consequently, the predictive capability of current models is limited, as it is profoundly undermined by both above-mentioned issues.

Because complex network theory and nonlinear time series analysis are generally considered domains of complex system science, the adoption of complex network methods has become a popular way of nonlinear time series analysis, thereby allowing fundamental questions regarding long-term dependencies and the chaotic property in time series prediction to be addressed \citep{pei2020geom, wang2012visibility, zou2019complex}.
For example, protein structural classes have been successfully predicted by mapping protein time series into recurrent networks \citep{olyaee2016predicting}.
Moreover, the virtual graphs derived from response time series of a marine system have helped forecast system catastrophes \citep{zhang2018predicting}, suggesting that the topological characteristics of time series graphs do contain latent information for regarding the future states of a chaotic trajectory.
For financial time series, it has been found that an exchange rate series converts into scale-free and hierarchically structured graphs \citep{yang2009visibility}.
Therefore, in the context of various graphs that are converted from time series, structural information, referring to the associations among temporal points and the node weights, provides promising assistance for financial time series prediction.
First, owing to the existence of explicit edges among distant nodes in converted graphs, the long-term dependencies in a time series can be directly captured through the associations among temporal points \citep{pei2020geom}. By bridging distant temporal points, long-range information can be delivered more quickly through these edges, thereby preventing information vanishing in recurrent deep learning.
Additionally, by identifying prominent content from chaotic time series, the weights of graph nodes can provide additional knowledge for temporal attention to tackle the chaotic property of financial time series \citep{wang2012visibility}.

In this paper, we propose a novel graph-based framework for stock prediction. Our framework consists of two main modules. The first is a time series embedding module, which is used to map time series into graphs and extract structural information from the corresponding graphs. In this module, given financial time series, which includes not only four kinds of stock prices (the closing price, high price, low price, opening price) but also the volume and amount of share trading at each time interval \citep{taylor2008modelling}, the visibility graph (VG) algorithm is employed for time series transformation \citep{lacasa2008time}, in which the mapped time series graph is denoted as the price graph. To overcome the above fundamental questions regarding prediction (e.g., long-term dependencies and the chaotic property), struc2vec is further adopted to learn node embeddings to preserve the associations among temporal points \citep{ribeiro2017struc2vec}, and the collective influence (CI) algorithm is employed for measuring the node weights of price graphs \citep{morone2015influence}. The second is a prediction module based on neural networks. Specifically, 
to complement the loss of temporal sequences in the structural information of stock time series, attention-based recurrent neural networks (RNNs) are employed to resolve the implicit long-term dependencies and chaotic evolution of the temporal points. Then, a self-attention layer is used to fully model the structure among stocks. Finally, with the obtained hidden representation of each stock, the corresponding movement direction is extracted from a final nonlinear fully connected layer.

To inspect the power of our framework, in this paper, we conduct numerous experiments on real-world market data from the Chinese market index (China Securities Index 300, i.e., CSI-300), which contains 300 stocks with data from 2010 to 2019. The experiment results demonstrate the effectiveness of the structural information extracted from price graphs for the task of stock prediction. 
A trading simulation is also performed based on signals produced by the prediction module to validate the profitability and stability of our framework.
The contributions of our work can be summarized as follows:
\begin{itemize}
\item For the first time, by utilizing complex network methods that bridge time series and graphs, we leverage the structural information obtained from market price data for stock prediction.
\item We develop a novel framework based on the structural information embedded in price graphs, and this framework is capable of addressing fundamental questions regarding long-term dependencies and the chaotic property in stock prediction.
\item We empirically reveal the effectiveness of structural information and the proposed framework for stock prediction on real-world data, i.e., our approach outperforms state-of-the-art baselines in terms of testing accuracy and obtains the highest average return (47.91\%) in trading simulations.
\end{itemize}

The remainder of this paper is organized as follows. Section \ref{sec:related_work} discusses the related literature. Section \ref{sec:framework} specifies the details of our proposed framework and structural information extraction methodology. Section \ref{sec:experiment_setup} depicts the experimental settings, including the data, compared baselines and model parameters. The subsequent section reveals the experimental prediction results.
We conduct a further analysis and market trading simulation in Section \ref{sec:discussion}. Finally, Section \ref{sec:conclusion} concludes this work and presents some limitations about future research directions.

\section{Related Work}
\label{sec:related_work}
In this section, we review the relevant literature streams regarding financial series prediction and graph learning to position our research vis-à-vis the findings from extant research.

\subsection{Financial time series prediction}
Before deep learning methodologies became popular, statistical and machine learning models were universally adopted for financial time series prediction because of their good interpretation capabilities.
Commonly used statistical models include the autoregressive moving average (ARMA) \citep{whittle1951hypothesis}, generalized autoregressive conditional heteroskedasticity (GARCH) \citep{bauwens2006multivariate} and nonlinear autoregressive exogenous (NARX) \citep{lin1996learning} models, which were employed to verify assumptions regarding the finance market, and predictions were accordingly based on these verified assumptions.
However, chaotic behaviors often appear with financial time series. As a result, the predictive capability of the models mentioned above is undermined because of their inability to shape the evolutionary process in financial systems \citep{farmer1987predicting}.
To improve the modeling of chaotic time series, a bunch of nonlinear learning models have been developed. In particular, Machine learning methods provide a strong capability to learn the underlying relationships among patterns between features and targets \citep{valiant1984theory, wu2020predicting}.
These methods focus on maximizing prediction accuracy based on a wide variety of models.
However, a limitation exists among these methods, that is the adoption of a predefined nonlinear framework which may not be consistent with the true underlying nonlinear form \citep{qin2017dual}.
In addition, the effectiveness of traditional statistical models or machine learning algorithms mostly relies on the quality of the input features, which enables improper features to possibly undermine model performance because of the chaotic property of financial time series.

In recent years, extensive studies have been undertaken to solve financial time series prediction problems using deep learning methods \citep{qin2017dual, fawaz2019deep, hao2020new} because of their powerful expression ability \citep{hornik1989multilayer}.
RNNs \citep{rumelhart1986learning}, deep neural networks specifically developed for sequence data, have a great vogue because of their superior performance in capturing nonlinear relationship.
However, traditional RNNs is insufficient in long-term dependencies capturing owing to the issue of gradients vanishing \citep{bengio1994learning}.
Based on ``memory cells'' that are designed to preserve information for a longer time, a special type of RNN, Long short-term memory (LSTM), has been developed and proven to be useful in predicting stock returns \citep{hochreiter1997long, fawaz2019deep}. Therefore, LSTM is constantly employed for sequential data or financial time series prediction and performs better than RNNs \citep{fawaz2019deep}. However, the shortage in capturing long-term dependencies still exists; that is, the performance of LSTM networks deteriorates rapidly when increases the length of the input sequence \citep{qin2017dual}. Furthermore, owing to the noise amplification in the model recurrence process, the chaotic property of financial time series worsens the prediction performance of the model \citep{hochreiter1997long, bayer2014fast}.

Because the concept of attention represents the human intuition by which some portions of data are given more emphasis than others, deep learning methods based on attention mechanisms are widely used to learn the complex dependencies among features in time series tasks.
Based upon this, a dual-stage attention-based RNN (DARNN) was proposed \citep{qin2017dual}. In the first encoder stage, an input attention mechanism automatically extracts the crucial input features at each iteration based on the previous hidden state of encoder, which gives greater emphasis to more informative features from chaotic series.
In the second decoder stage, a temporal attention mechanism is designed to select crucial encoder hidden states by referring to all time steps, thereby rebalancing the information at each temporal point to capture long-term dependencies.
In another recent work, a cross-attention stabilized fully convolutional neural network (CA-SFCN), which also adopts variable and temporal attention, was proposed to classify multivariate time series \citep{hao2020new}.
While these models implicitly capture certain long-term dependencies, there is still rare attention that explicitly exploits the structure within time series, i.e., the direct links among distant temporal points. Through the bridging of distant temporal points to form edges, long-range information could be directly obtained through these edges to prevent information vanishing in traditional deep learning.
We believe that more accurate predictions can be obtained by capturing this explicit structural information of the given time series.

\subsection{Time series graphs}
With increasing literature leveraging complex network methods for characterizing dynamical systems derived from time series, the adopting of complex network methods has become an active realm of nonlinear time series analysis, which has provided a guideline for addressing fundamental questions regarding long-term dependencies and the chaotic property in time series prediction \citep{pei2020geom, wang2012visibility, zou2019complex}.
By transforming time series into graph, researchers are capable of measuring the structural properties of time series and capturing the hidden structures embedded within chaotic temporal points.
Based on different rationals, there are three main kinds of methods designed to map a time series into a graph: (1) recurrence networks (RNs), which emphasizes the mutual statistical similarities or metric proximities among various segments of the time series \citep{donner2010recurrence}; (2) VGs, which depict local convexities or record-breaking properties within a signal time series \citep{lacasa2008time}, and (3) transition networks (TNs), which profiles the transition probabilities between discrete states \citep{nicolis2005dynamical}. Among these three complex methods, there are many parameters need to be optimized in RNs and TNs when converting time series. Although, there is also literature discussing the optimal choice about these parameters of RNs and TNs, a simple algorithm (VG) is much preferred for financial time series analysis because of the parameter-free nature \citep{zou2019complex}. In particular, the associated VG refines certain important features from the original time series, i.e., the structure within the time series \citep{zou2019complex}.

Chaotic time series widely exist in various scenarios, including finance, biology, and meteorology, and there has been a wide range of recent applications of complex network methods for such chaotic data \citep{olyaee2016predicting, zhang2018predicting, yang2009visibility}.
As reported in \cite{olyaee2016predicting}, the structural classes of proteins have been successfully predicted by transforming protein time series into recurrence networks.
Moreover, another recent study showed that the virtual graph derived from response time series of a marine system can be used to forecast system catastrophes \citep{zhang2018predicting}.
While in financial studies, the degree distributions of mapped graphs derived from the growth rates of gross domestic product series were found to be scale-free, and the degree distributions of converted graphs derived from the growth rates of three industry series are almost exponential \citep{yang2009visibility, wang2012visibility}.
Another study found that the markets in developed countries differ significantly from developing countries through analyzing the time series graphs mapped from stock market \citep{cao2014unraveling}.
Specifically, the complexity of developing markets is disturbed and relatively low over some periods while more stable and stronger over time for mature stock markets, suggesting a stronger long-range price memory and indicating that transforming financial time series into graphs can effectively capture the chaotic characteristics of stocks.

Therefore, in the context of various graphs converted from time series, previous complex network methods naturally provided promising assistance for financial time series prediction.
First, owing to the existence of explicit edges among distant nodes in converted graphs, the long-term dependencies in time series graphs can be directly captured through the associations among temporal points. Additionally, by identifying prominent content from chaotic time series, the weights of graph nodes provide additional knowledge for learning temporal attention to tackle the chaotic property of financial time series.
However, despite the great potential of structural information for financial time series forecasting, little attention has been paid to the employment of complex network methods for obtaining more accurate predictions. Moreover, how to integrate the structural information of time series networks and deep learning methods remains an open problem.

\subsection{Graph embedding}
To employ graph structures for deep learning, graph embedding has aroused much research interest.
With the advantage of preserving node content, graph structure, and additional information, graph embedding is capable of embedding graph nodes into latent, low-dimensional spaces \citep{zhang2018network}.
After obtaining the new node representations, conventional vector-based learning methods can be conveniently and efficiently employed for graph analysis tasks; this inspires us to take advantage of structural information for time series prediction.

Traditional efforts in learning low-dimensional vectors for vertices in networks have been considerably successful with respect to performing prediction and classification tasks \citep{ribeiro2017struc2vec, grover2016node2vec, dai2021attention, yu2020estimating}. Based on the proximity of nodes derived from learned embeddings, the future interactions between users can be extracted from a social network \citep{song2009scalable}. In addition, through dynamically computing a user's recent preferences by referring to the embedding of check-in points of interest (POIs), a location-based model was introduced to help discover attractive and interesting POIs \citep{xie2016learning, lin2021go}. For financial problems, with a bipartite network constructed from mutual fund shareholding data in the real world, the intrinsic properties of stocks were extracted from the latent space to optimize technical indicators and target the critical factors of market crashes \citep{li2019individualized}. Besides the interrelationships among stocks, various networks, which consists of financial institutions, cryptocurrencies, stock indices and stock sectors, are constructed for investment/portfolio assistances \citep{bouri2018does, ji2018network, shahzad2021impact}. While excellent performance has been confirmed, these target graphs may not be applicable to more refined problems, such as daily stock prediction, because these graphs contain only coarse-grained information.
Moreover, due to the connection among nodes established on assets, the weaved graphs are time invariant or evolve with a low frequency (e.g., monthly or seasonally in general); that is, the embedded information cannot change in a timely manner with the dynamics of the whole market, which may result in lower effectiveness in terms of daily stock prediction. Therefore, these target graphs are often adopted to identify the market states over a period or assist other methods which are capable of timely reaction \citep{shahzad2021extreme, bouri2018does, shahzad2021impact}. With the emergence of time series graphs, more refined structural information can be obtained, which suggests great potential for financial time series prediction.

The extant literature on complex networks and graph embedding implies that the structural information extracted from time series graphs is capable of tackling issues regarding long-term dependencies and the chaotic property. In this paper, based on this structural information obtained from time series graphs, we propose a new framework to obtain more accurate stock predictions.

\begin{figure}[!htbp]
  \centering
  \includegraphics[width=1.\textwidth]{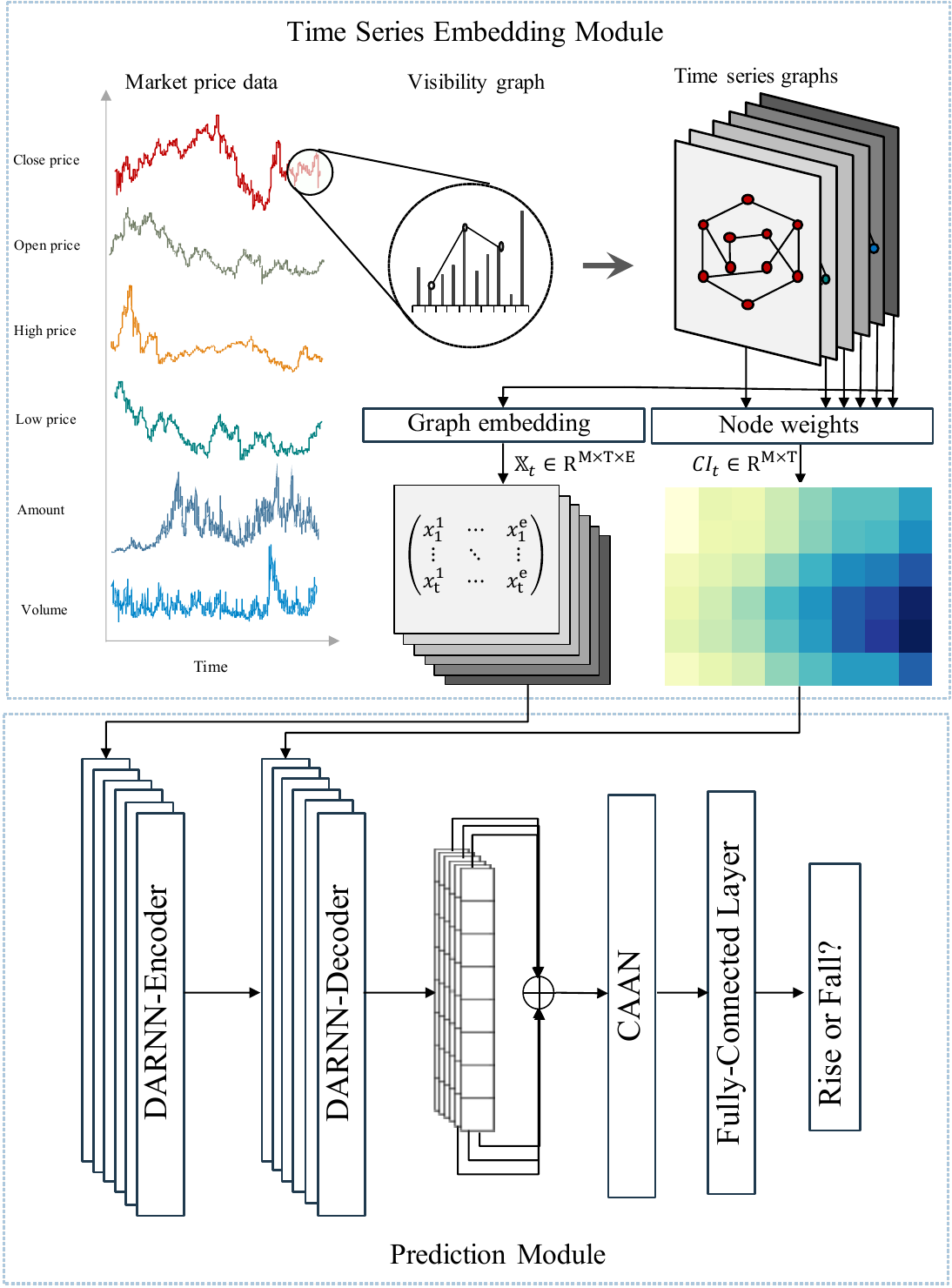} 
  \caption{\textbf{Computational flow of the proposed framework.} Let $P_t\in\mathbb{R}^{M\times T}$ denote the history price data for a stock at time $t$, the node vectors after embedding is $\mathbb{X}_t\in \mathbb{R}^{M\times T \times E}$ and node weights is $CI_t\in \mathbb{R}^{M \times T}$, where $M$ is number of stock quote data (six here), $T$ is the length of the lookback window and $E$ is the embedding size of graph nodes.}
  \label{fig:frameword_overview} 
\end{figure}

\section{Proposed framework}
\label{sec:framework}
This paper presents a novel trend prediction framework for stock prices. To conquer the shortcomings of the financial time series forecasting methods mentioned above, a module based on complex network methods and graph embedding is introduced to extract structural information from mapped graphs that structurally connect distant price points.
The predictions of stock trends are obtained from several attention-based layers and a fully connected classification layer.
Figure \ref{fig:frameword_overview} illustrates the proposed framework, which is constructed with two modules.
The first module is a time series embedding module, which takes raw market price data as inputs and aims to extract structural information, referring to the associations among temporal points and the node weights.
The second module is a prediction module. Based on deep learning methods, stock representations are learned from structural information by incorporating temporal sequences for stock trend forecasting.
The different modules of the proposed framework are discussed below in turn.

\subsection{Time series embedding module}
In this module, we convert raw market price data into time series graphs and measure the weights of graph nodes. Second, we explore the topological properties of the constructed graphs associated with raw market data, namely, the structural representations.

\subsubsection{Time series graphs}
As discussed above, there are three main types of complex network approaches for mapping individual time series into graphs, i.e., RNs, VGs and TNs. Among these three complex methods, the VG algorithm is widely employed for financial time series analysis because it is not influenced by any algorithmic parameters and maps time series into scale-free graphs \citep{lacasa2008time}. Thus, we take the VG algorithm to map raw market price data into time series graphs, i.e., price graphs.

\begin{figure}[!htbp]
  \centering 
  \subfigure[Visibility graph]{ 
    \includegraphics[width=.55\textwidth]{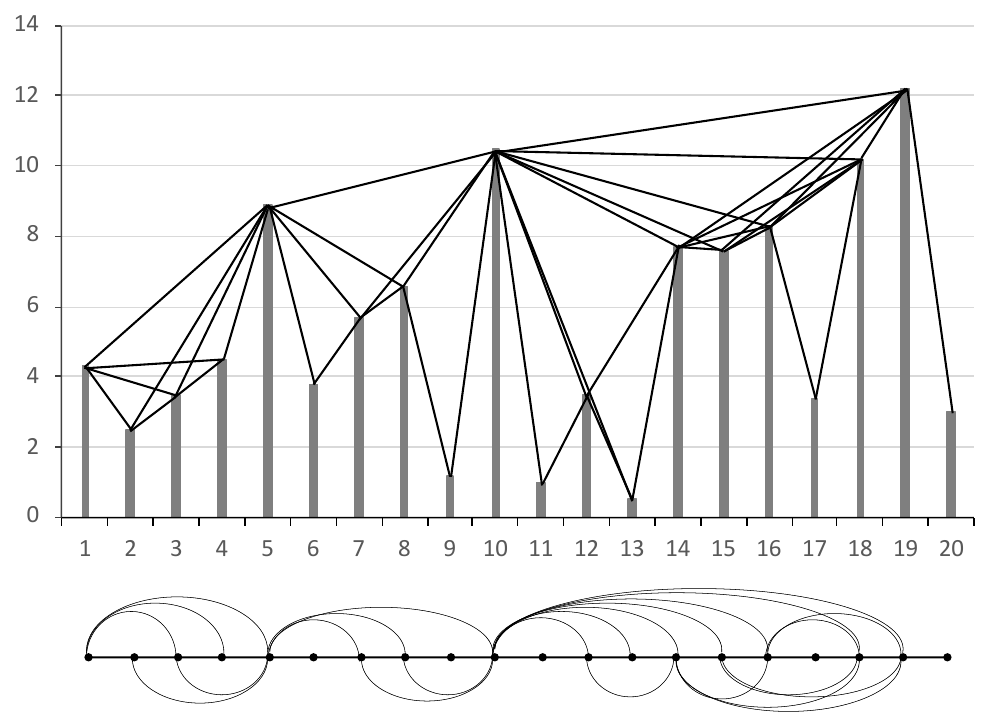} 
    \label{fig:vg_sample} 
  } 
  \subfigure[Time series graph with node weights]{ 
    \includegraphics[width=.4\textwidth]{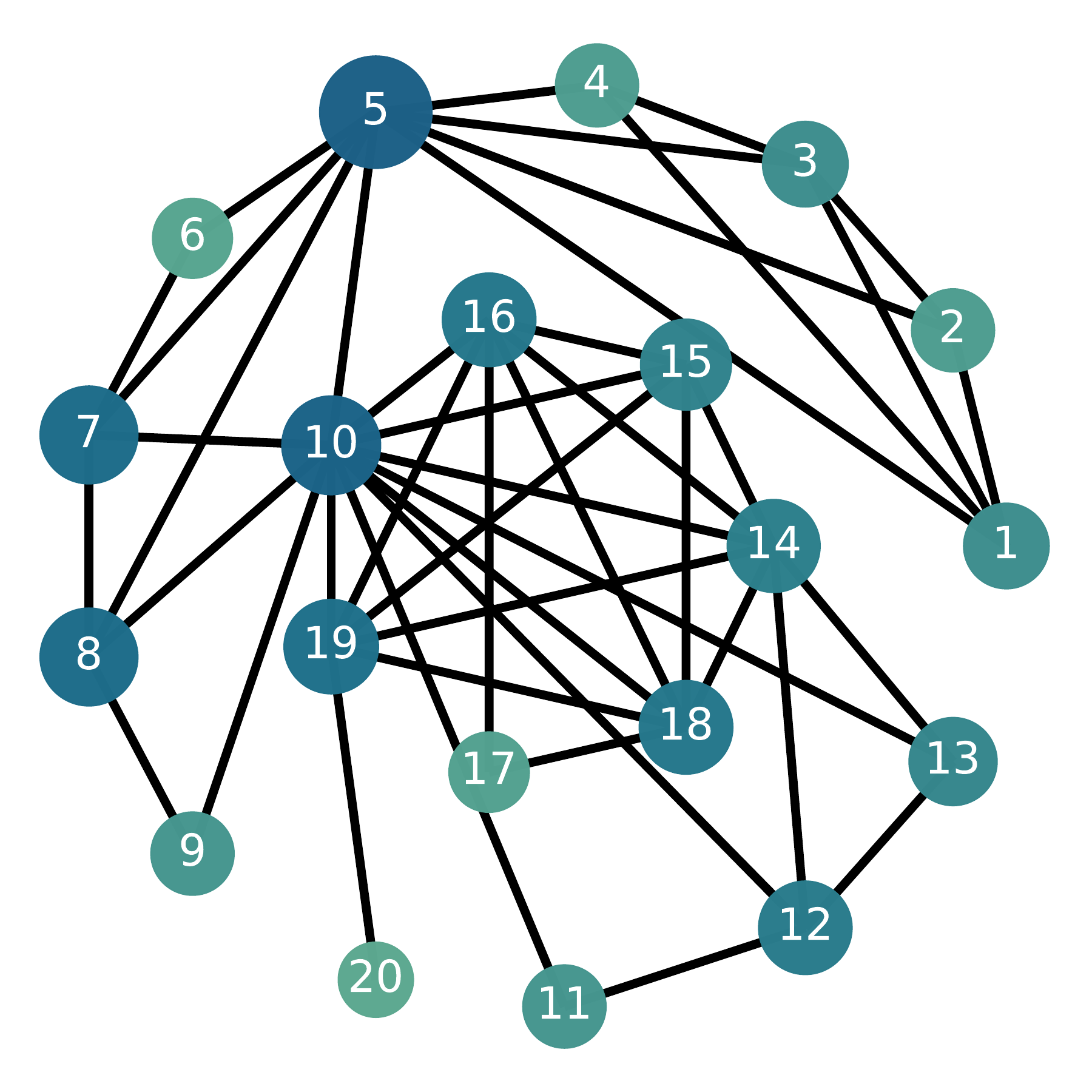} 
    \label{fig:vg_ci_sample} 
  } 
  \caption{\textbf{Instance of a time series containing 20 temporal points and the associated VG derived from the VG algorithm.}}
\end{figure}

Consider a stock price series $p_t$ with a length of $T$ at time $t$. 
Each vertex in the converted graph corresponds to data point in original series. Based on the principle that if two data points can mutually be seen in the bar chart of corresponding time series, an edge between the two points is established; put differently, two temporal points in series can be connected by a straight line when there are not any intermediate data heights intersect this ``visibility line''.
In a formal manner, a VG can be transformed from a time series under the next visibility specification \cite{lacasa2008time}: given two data points ($t_i$, $p_i$) and ($t_j$, $p_j$) where $p_i$, $p_j\ >\ 0$ in a time series, there is a visibility line that connects the two data points in converted graph if and only if all data points ($t_k$, $p_k$) such that $t_i<t_k<t_j$ satisfies
\begin{equation}
    p_k < p_i + \frac{t_k-t_i}{t_j-t_i}(p_j-p_i).
    \label{eq:vg_algorithm}
\end{equation}
For graphs converted from stock prices, each vertex represents the daily price under real-world trading circumstances. We can view the graph edges as signals of no abrupt price changes occurring during the period between two temporal points.
Furthermore, on the basis of Equation \ref{eq:vg_algorithm}, we can find that the shortest path between any two temporal points in the converted graph is definitely shorter than their original time interval, which indicates that time series graphs are adept at capturing long-term dependencies because the information embedded in early prices can be sooner obtained by later points without long-range transfers and information vanishing.

VGs are connected graphs and the construction process is invariant even if a series of basic transformations is performed on the original series, such as vertical and horizontal translations \citep{lacasa2008time}.
An example of VG algorithm can be seen in Figure \ref{fig:vg_sample}, where we present a converted VG and the original time series with 20 data points. As seen, temporal points are connected by edges as long as they can see each other.
Furthermore, we are capable of measuring the weight of every temporal point for chaotic property settling after we obtain the graph structure within the time series. In traditional complex network analysis, there are many metrics used to identify vital nodes, such as degrees, k-cores, and cluster coefficients. In this study, the collective influence (CI) algorithm is selected due to its low computational consumption yet excellent performance in characterizing the node influence with regard to influence as a collective attribute, instead of a local feature like the node's degree \citep{morone2015influence}.
In the CI algorithm, there is a $Ball(i,l)$, which contains a bunch of nodes within a ball of radius $l$ (the length of shortest path) around node $v_i$. The frontier of this ball is defined as $\partial Ball(i, l)$. Then, the CI index of node $v_i$ of radius $l$ is
\begin{equation}
    CI_l(i) = (d_i-1)\sum_{j\in \partial Ball(i, l)}(d_j-1),
\end{equation}
where $d_i$ is the degree of node $v_i$ and $l$ is a nonnegative integer and given according to the graph diameter (normally less than diameter) \footnote{In general, $l$ was set to 3 and achieved sufficient performance in \citep{morone2015influence}, while we assign 2 to $l$ because our mapped graphs with 20 nodes are far smaller than the graphs in \citep{morone2015influence}}. A time series graph with CI as the node weight converted from the time series in Figure \ref{fig:vg_sample} is shown in Figure \ref{fig:vg_ci_sample}. In this figure, nodes with higher weights are marked with deeper colors and greater sizes.

In this phase, we use the vector $P_t$ to denote the historic state of a stock at time $t$, where $P_t \in \mathbb{R}^{6\times T}$ and consists of the raw market price data (the closing price, high price, low price, opening price, amount and volume), and $T$ is the length of the lookback window of $t$. Through the VG algorithm, converted price graphs $\mathbb{G}_t$ are obtained, where $\mathbb{G}_t = [G^C_t, G^O_t, G^H_t, G^l_t, G^A_t, G^V_t]$ and each graph has $T$ nodes. Furthermore, before extracting structural information, we measure the node weights $CI_t$ of the converted graphs through the CI algorithm, where $CI_t \in \mathbb{R}^{6 \times T}$.

\subsubsection{Graph embedding}

In this paper, graph embedding based on vector representations is further employed, as this not only complements the loss of mapped price graphs in the temporal sequence but also incorporates structural information into deep learning methods.
To retain the structural properties of graphs regarding associations among their temporal points, we adopt struc2vec \citep{ribeiro2017struc2vec} in this paper.
Struc2vec is a skip gram \citep{mikolov2013efficient}-based graph embedding algorithm that aims to learn a mapping $g: v \in V \mapsto \mathbb{R}^{\left | V\right | \times d}$.
According to the structural similarities based on nodes' $k$-hop neighborhoods, struc2vec is able to learn a vector-based representations that capture the structural roles of the nodes.
Specifically, the method involves executing numerous random walks over the graph from each node. The co-occurrences of nodes in a short window are captured based on the sequences of these walks, which can be used to tackle the diffusion in the neighborhood around every node in the graph and explore the local topology structure around a vertex.
The embedding method is designed to learn a representation that enables the estimation of the possibility of a node $u$ showing together with other nodes in the subwindow of a short random walk:
\begin{equation}
    \underset g{\mathrm{max}} \sum_{u\in V} \mathrm{log} Pr(N(u)|g(u)),
\end{equation}
where $N(u)$ represents the neighboring nodes of node $u$. The likelihood of a vertex $v$ appearing together with $u$ is estimated by using a softmax function:
\begin{equation}
    Pr(v|u) = \frac{exp(g(v)\cdot g(u))}{\sum_{v_i\in V}exp(g(v_i) \cdot g(u))}.
\end{equation}
Furthermore, to measure the structural similarities between nodes, struc2vec \citep{ribeiro2017struc2vec} generates a series of weighted assistant graphs ${\boldsymbol{g}}_k$, $k=\{1, 2,\dots,k^*\}$ that derived from original graph, in which the assistant graphs capture the structural similarities between nodes' $k$-hop neighborhoods, and $k^*$ is the diameter of original graph. More concretely, each assistant graph is a weighted undirected complete graph. As for the weights of edges, $R_k(v)$ is defined as the ordered degree sequence with nodes that are precisely $k$ hops from $v$, and $w_k(v, u)$ that measures the edge weights in an assistant graph ${\boldsymbol{g}}_k$ is recursively defined as
\begin{equation}
    w_k(v, u) = w_{k-1}(v, u) + d(R_k(v), R_k(u)),
\end{equation}
where $w_0(v, u)$=$0$ and $d(R_k(v), R_k(u))$ measures the difference between $R_k(v)$ and $R_k(u)$. 
And struc2vec can generate vertex sequences based on the edge weights in these weighted assistant graphs.
Then, to maximize the probability of neighboring nodes in the local area appearing together with the central node, a neural network is trained by using the skip gram architecture. Finally, the output of the hidden layer of trained neural network is obtained as the embedding of nodes. Through embedding price graphs into vector representations, struc2vec further enriches the information of time points by considering the global dependencies in representations and enhances the description of raw time series. The long-term dependencies among temporal points can be captured owing to the feature of struc2vec in structural similarity measuring.

In this phase, with the converted price graphs $\mathbb{G}_t$, we obtain the tensor representations $\mathbb{X}_t$ for all graphs through struc2vec at time $t$, where $\mathbb{X}_t = [X^C_t, X^O_t, X^H_t, X^L_t, X^A_t, X^V_t]$, $X_t \in \mathbb{R}^{T \times E}$ and $E$ is the embedding size.

\subsection{Prediction module}
\begin{figure}[!htbp]
  \centering
  \includegraphics[width=1.\textwidth]{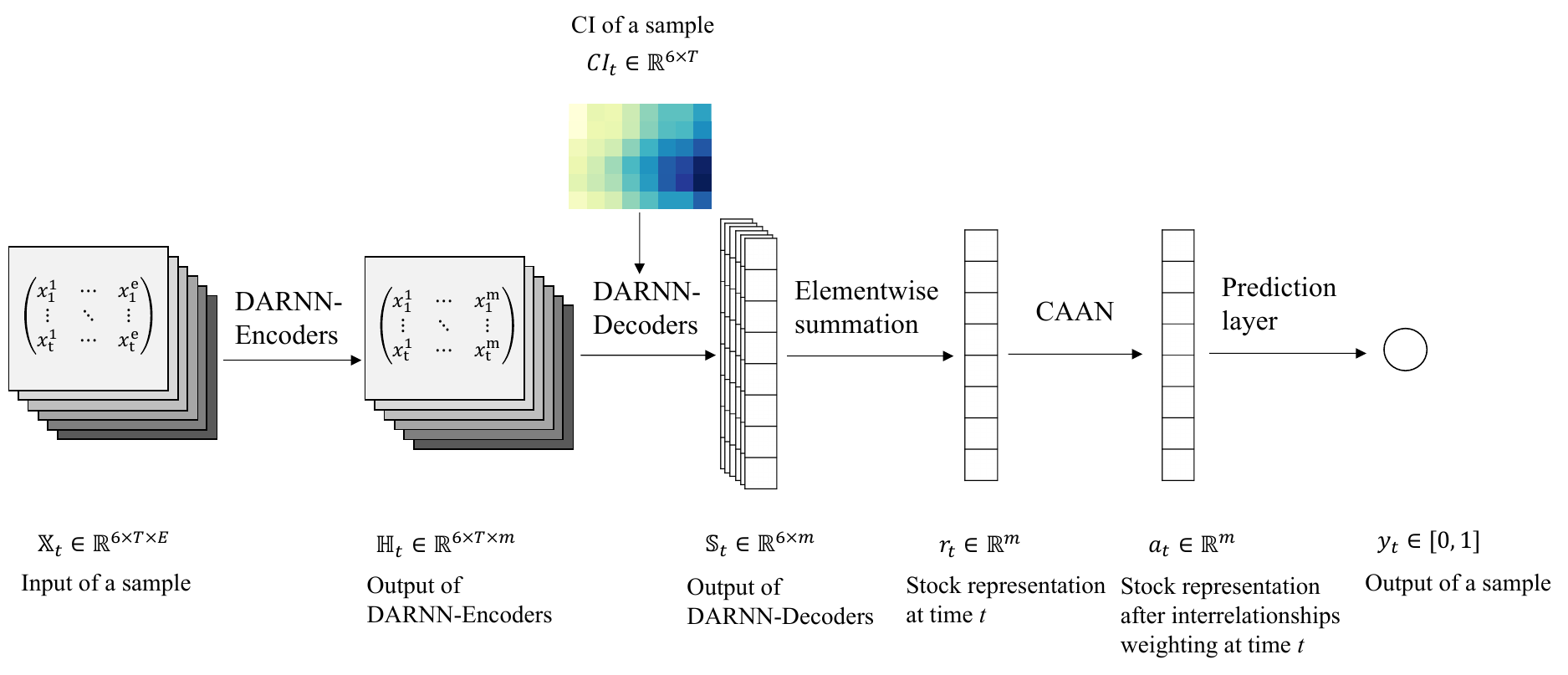} 
  \caption{\textbf{Computational flow of the prediction module.} Here, we take a sample at time $t$ for example. In which $T$ is the lookback window of history prices at time $t$. $E$ is the embedding size of struc2vec for each graph node. In experiments, we set hidden vectors of our deep learning method to a fixed size $m$.} 
  \label{fig:prediction_data_flow} 
\end{figure}

In this section, we introduce the classification module, which is constructed with two main parts. The first key part has several (e.g., six here) DARNNs with node weights for the temporal attention network \citep{qin2017dual}. The goal of the first part is to automatically extract input series representations by incorporating temporal information and structural information \footnote{Temporal information is obtained through organizing the model inputs in the order of time sequences.}. Based on this model, a stock representation $r_t$ is obtained from the combination of several DARNNs' outputs for each stock at time $t$.
The second key part is a cross-asset attention network (CAAN), which is used to describe the interrelationships among the stocks \citep{wang2019alphastock}. Finally, the classification results are given by a fully connected classification layer. And to further illustrate the detailed data transformation in the prediction module, Figure \ref{fig:prediction_data_flow} shows the input and output for each step \footnote{We omit the presentation of historical price input of DARNN-Decoder to emphasize the structural information input.}.

\subsubsection{Structural information learner}
The structural information learner contains multiple DARNNs that incorporate node weights.
DARNNs \citep{qin2017dual} are employed as our essential components not only because of their capability of selecting crucial variables and temporal points but also because of their excellent performance regarding time series forecasting in comparison to LSTM \citep{hochreiter1997long} and attention-based LSTM \citep{bahdanau2015neural}. Without the loss of any generality, we show the learning process in one of the six DARNN models for illustration. Each DARNN is an encoder-decoder network and have the same learning process with different inputs. Specifically, the encoder is basically an LSTM that learns the input sequences as a hidden representation that considers input attention. For time series prediction, letting $X=(x_1, x_2, \cdot \cdot \cdot , x_T)\in\mathbb{R}^{T\times E}$ denote the input sequence, where $x_t \in \mathbb{R}^E$ and $E$ is the graph embedding size, the encoder is used to recursively learn a hidden vector from $X$:
\begin{equation}
    h_t=\mathrm{LSTM}(h_{t-1}, x_t), t\in[1, T],
\end{equation}
where $h_t\in \mathbb{R}^m$ is the hidden vector encoded by LSTM at time step $t$ and $m$ is the size of each hidden vector.

As attention mechanisms are widely employed in deep neural networks, a DARNN integrates the input attention into the encoder stage, which can automatically select the appropriate input series to give more emphasis to informative features obtained from chaotic series.
Letting $x^k = (x^k_1, x^k_2, \cdot \cdot \cdot, x^k_T)^\top \in \mathbb{R}^T$ denote the $k$-th input series, the input attention is calculated through a deterministic attention model, which adopts the hidden state $h_{t-1}$ and cell state $s_{t-1}$ of the encoder LSTM unit with
\begin{equation}
    c^k_t = v^\top_c \mathrm{tanh}(W_c[h_{t-1};s_{t-1}] + U_cx^k)
\end{equation}
and
\begin{equation}
    \alpha^k_t = \frac{\mathrm{exp}(c^k_t)}{\sum^n_{i=1}{\mathrm{exp}(c^i_t)}},
\end{equation}
where $v_c\in \mathbb{R}^T$, $W_c \in \mathbb{R}^{T\times 2m}$ and $U_c\in \mathbb{R}^{T\times T}$ are learnable parameters. At time step $t$, the importance regarding the $k$-th input series is estimated by $\alpha^k_t$. To normalize the attention weights, a softmax is applied to $c^k_t$.
Thus, the updating of input series with attention weights can be formulated as
\begin{equation}
    \tilde{x}_t = (\alpha^1_tx^1_t, \alpha^2_tx^2_t, \cdot \cdot \cdot, \alpha^n_tx^n_t).
\end{equation}
Then the updating of the hidden state at time $t$ is
\begin{equation}
    h_t = \mathrm{LSTM}(h_{t-1}, \tilde{x}_t).
\end{equation}
By measuring the attention weight of each input series, the encoder can adaptively select crucial series rather than assuming all the input series to be equal. Taking the obtained structural information $\mathbb{X}_t$ as input, hidden features $\mathbb{H}_t = [H^C_t, H^O_t, H^H_t, H^L_t, H^A_t, H^V_t] \in \mathbb{R}^{6\times T \times m}$ for six prices can be obtained from the DARNN-Encoders.

Together with the encoder that applies the input attention, a decoder that incorporates temporal attention is also proposed to automatically select appropriate hidden vectors of encoder from among all time intervals, as these can implicitly and appropriately capture the long-term dependencies within a time series. More concretely, the temporal attention among encoder hidden states is also calculated through a deterministic attention model, which adopts the hidden state $h'_{t-1} \in \mathbb{R}^m$ and cell state $s'_{t-1} \in \mathbb{R}^m$ of the decoder LSTM unit with
\begin{equation}
    d^i_t = v^\top_d \mathrm{tanh}(W_d[h'_{t-1};s'_{t-1}]+U_dh_i),\ 1\leq i\leq T
    \label{temporal_tanh}
\end{equation}
and
\begin{equation}
    \beta^i_t = \frac{\mathrm{exp}(d^i_t)}{\sum^T_{j=1}\mathrm{exp}(d^j_t)},    
    \label{temporal_softmax}
\end{equation}
where $U_d\in\mathbb{R}^{m\times m}$, $W_d\in\mathbb{R}^{m\times2m}$ and $v_d\in\mathbb{R}^m$ are parameters to be learned and $m$ is the size of each hidden state.
Similarly, $\beta^i_t$ is a temporal attention weight that measures the importance regarding the $i$-th encoder hidden state, thereby rebalancing the information at each temporal point to capture long-term dependencies and filter out chaotic temporal points.
A softmax is also applied to $d^i_t$ to normalize the attention weights.
Thus, with the obtained temporal attention, all encoder hidden states $\{h_1, h_2, \cdot \cdot \cdot, h_T\}$ are weightedly summed as a context vector $e_t$:
\begin{equation}
    e_t = \sum^T_{i=1}\beta^i_t h_i.
\end{equation}
The context vector $e_t$ varies for each time step.

In this basic decoder, the temporal weights modeled by Equations \ref{temporal_tanh} and \ref{temporal_softmax} are directly learned from the hidden encoder representation.
In fact, as we discussed in the previous section, we can use priori knowledge to enhance the temporal weight learning process.
Given the previously mentioned node weights of the price graphs, we use the node weights as the knowledge-based attention over the temporal points to highlight informative points and address the chaotic property of price series.
Then, for each hidden state $h_i$, we update the attention weight $d^i_t$ as
\begin{equation}
    \tilde{d}^i_t = v^\top_d \mathrm{tanh}(W_d[h'_{t-1};s'_{t-1}]+U_dh_i+w_dCI_i),\ 1\leq i\leq T, 
    \label{eq:decoder_attn_ci}
\end{equation}
where $w_d$ is a learnable parameter and $CI_i$ denotes the node weight of temporal point $i$. In this way, the relative importance levels of temporal points in the price graphs are introduced as weights to enhance attention learning and chaotic property settling. The nodes that obtain critical positions in the time series contribute more to the final sample representations and make more accurate predictions. After reweighing the temporal attention, the combination of the updated context vector $\tilde{e}_t$ with the stock price series $\{p_1, p_2, \dots, p_{T-1}\}$ is denoted as $\tilde{p}_{t-1} = w_p[p_{t-1};\tilde{e}_{t-1}]+b_p$, where $[p_{t-1};\tilde{e}_{t-1}]\in \mathbb{R}^{m+1}$ is a concatenation of history price and context vector, $w_p \in \mathbb{R}^{m+1}\ and\ b_p\in\mathbb{R}$ are parameters to learn. Accordingly, the newly $\tilde{p}_{t-1}$ is adopted to update the hidden feature of LSTM in decoder as $h'_t = \text{LSTM}(h'_{t-1}, \tilde{p}_{t-1})$. Finally, the output of DARNN-Decoder can be formulated as 
\begin{equation}
S = \tilde{w}[h'_T; \tilde{e}_T]+\tilde{b},
\end{equation}
where $S \in \mathbb{R}^m$ and $[h'_T; \tilde{e}_T]\in\mathbb{R}^{m+m}$ is a concatenation of the hidden feature $h'$ and and the updated context vector $\tilde{e}$ at the last step of LSTM. The parameters $\tilde{w} \in\mathbb{R}^{m\times 2m}$ and $\tilde{b}\in\mathbb{R}^m$ map the concatenation to the size of the decoder hidden states.

With the hidden features $\mathbb{H}_t$ learned from DARNN-Encoder, the stock representations $\mathbb{S}_t=[S^C_t, S^O_t, S^H_t, S^L_t, S^A_t, S^V_t]$ for six price series at time $t$ are computed through several DARNNs, where $S_t \in \mathbb{R}^m$ and $m$ is the hidden size of each DARNN's decoder. Then, with an elementwise summation layer, six representations are merged into one $r_t=\sum_{I\in\{C,O,H,L,A,V\}}S^I_t \in \mathbb{R}^m$ as the stock representation at time $t$.

\subsubsection{The CAAN model}
While acquiring the stock representations, we adopt a CAAN based on self-attention to utilize the interrelationships among stocks. 
Specifically, based on the merged stock representation $r_i$ \footnote{The omittance of time $t$ does not cause the loss of any generality} of stock $i$, we calculate three vectors $q^i=W_qr^i$, $k^i=W_kr^i$ and $v^i=W_vr^i$ as the query vector, key vector and value vector, respectively, where $W_q$, $W_k$, and $W_v$ are learnable parameters. The interrelationships between stock $i$ and other stocks within a batch are computed by using the query vector $q^i$ of stock $i$ to query the key vectors of other stocks, i.e.,
\begin{equation}
    l^{ij} = \frac{q^{i\top} \cdot k^j}{\sqrt{D_k}}
\end{equation}
and
\begin{equation}
    \gamma^{ij} = \frac{\mathrm{exp}(l^{ij})}{\sum^I_{{j}'=1}\mathrm{exp}(l^{i{j}'})},
\end{equation}
where $D_k$ is a rescaling parameter set in line with \cite{vaswani2017attention} and $I$ is the size of each sample batch.
Afterwards, we use the interrelationships {$\gamma^{ij}$} to weighted the value vectors {$v_j$} into an attention representation
\begin{equation}
    a^i = \sum^I_{j=1}\gamma^{ij} \cdot v^j
\end{equation}
for stock $i$. Finally, to forecast the future price trend, the scalar prediction score $\hat{y}^i$ for stock $i$ is computed by a feed-forward layer and a sigmoid transformation:
\begin{equation}
    \hat{y}^i = \mathrm{sigmoid}(W_{fc}a^i + b_{fc})    
\end{equation}
where $W_{fc}$ and $b_{fc}$ are the linear parameter and the bias to be learned, respectively. Finally, based on the computation flow in Figure \ref{fig:prediction_data_flow}, the learning and optimization process of our prediction module are summarized in Algorithm \ref{code:prediction_module}.

\begin{algorithm}[!ht] 
\setstretch{1.35}
\caption{Optimization of our prediction module within a sample}
\label{code:prediction_module} 
{\bf Input:} input sample $\mathbb{X}\in\mathbb{R}^{6\times T\times E}$, corresponding $CI \in\mathbb{R}^{6\times T}$, historical price series $P\in \mathbb{R}^{6\times T}$, labeled target $y$ \\ 
{\bf Output:} predicted label $\tilde{y}$ of input
\begin{algorithmic}[1]
\STATE Denote all parameters of the prediction model as $\mathbb{W}$;
\STATE Initialize the parameters $\mathbb{W}$;
\FOR{$epoch\in [1,\dots, maxIteration]$} {
	\STATE $\mathbb{H} = \text{DARNN-Encoders}(\mathbb{X})$, where $\mathbb{H}\in\mathbb{R}^{6\times T \times m}$;
	\STATE $\mathbb{S} = \text{DARNN-Decoders}(\mathbb{H}, CI, P)$, where $\mathbb{S}\in\mathbb{R}^{6\times m}$;
	\STATE $r=\sum_{i=0}^{6}\mathbb{S}_i$, where  $r\in\mathbb{R}^m$;
	\STATE $a=\text{CAAN}(r)$, where $a\in\mathbb{R}$;
	\STATE $\tilde{y}=f_{prediction}(a)$, where $\tilde{y}\in\mathbb{R}$;
	\STATE // computing loss on a batch sample;
	\STATE $\mathcal{L} = \text{LossFunction}(y, \tilde{y})$;
	\STATE Update hidden parameters $\mathbb{W}$ with gradient decent ($\mathcal{L}|\mathbb{W}$);
}
\ENDFOR
\end{algorithmic} 
\end{algorithm}

\section{Experiments}
\label{sec:experiment_setup}
To evaluate the proposed framework, we use stock data from the China A-share market. In the following subsections, we introduce the details of the data and the compared methods.
We also depict the detailed experimental settings of the training and testing process for our proposed framework and baselines.

\subsection{Data}
We collect the daily quote data of CSI-300 component stocks from the China A-share market from January 1, 2010, to December 31, 2019 to cover comprehensive patterns in price trends and avoid the external shock from the COVID-19 on model validation. In particular, China stock market has great research value and plenty of research attention has been devoted to it \cite{xu2019interconnectedness, bouri2018does, ji2018network}, and in which the CSI-300 index selects the most liquid A-share stocks and aims to reflect the overall performance of the China A-share market \citep{long2019deep}.
In addition to the explicit influences from the supply and demand of investors, the price of a stock can also be changed owing to some firm operations, including stock splits, rights offerings and dividend payouts or distributions, and the prices need to be adjusted after these actions. In this case, the adjusted price is more insightful for investigating historical returns because it reflects an accurate performance of a firm's market value beyond the raw market trading prices.
The adjusted daily quote data are obtained from Tushare \footnote{http://tushare.org.}, an open source financial data package \citep{wu2020predicting}.
The data contain daily stock prices, including closing prices, low prices, high prices, opening prices, amounts and volumes.

\subsection{Baselines}
Due to the ubiquity of deep neural networks, most recent efforts regarding stock prediction have been devoted to leveraging LSTM \citep{nelson2017stock}, CNNs \citep{long2019deep, hao2020new} and attention-based approaches \citep{qin2017dual, hao2020new} with market price data as inputs. Competitive performance is also achieved by these novel methods.
Note that among these state-of-the-art prediction models, to the best of our knowledge, none of them leverage the structural information extracted from time series graphs.
To inspect the power of our framework, which investigates such important information, our proposed framework is compared to the following models:
\begin{itemize}
    \item LSTM: a basic LSTM network used to predict the future trends of stock prices based on historical price data \citep{nelson2017stock}.
    \item DARNN: a dual-stage attention-based RNN that employs input attention and temporal attention in the encoder and decoder stages, respectively \citep{qin2017dual}.
    \item DARNN-SA: an extension of the DARNN that employs a self-attention layer between the output of the DARNN and the prediction layer.
    \item MFNN: a multifilter deep learning model that integrates convolutional and recurrent neurons for feature extraction with respect to financial time series and stock prediction \citep{long2019deep}.
    \item CA-SFCN: a fully convolutional network (FCN) incorporating cross attention (CA), in which CA is also used as dual-stage attention for the variable and temporal dimensions (with temporal attention first) \citep{hao2020new}.
\end{itemize}
In particular, in line with their original model inputs, all five baseline methods take market price data as inputs. In addition, we further make the comparison between our proposed framework with the classical time series approaches, i.e., ARMA and GARCH models.

\subsection{Parameter settings}
To prevent data snooping, experiment data sets are strictly split according to the sample dates. For example, we use the data from Jan. 2010 to Dec. 2018 as the training and validation sets and the rest as the test set, which includes the whole year of 2019 \footnote{We also test our framework on CSI-300 across 2015 to further inspect its capability under extreme market environment, stable and competitive performances still can be obtained by our proposed framework.}.
During the training process, 70\% samples are randomly selected as the training set and the remaining 30\% of the samples as the validation set to take full advantage of the historical information.
For each stock, we take the historical information with a window size of 20 days \citep{qin2017dual, feng2019enhancing} to predict the price trend of the next day.
The target of our experiments is defined as follows:
\begin{equation}
    y = \left\{\begin{array}{ll}
        1\ (\uparrow)\,, & p^c_{t+1}>p^c_t \\
        0\ (\downarrow)\,, & \text{otherwise,}
        \end{array}
    \right.
\end{equation}
where $p^c_t$ is the stock closing price at time $t$.
Furthermore, in the experiment, the test period is set to a three-month sliding window; thus, 4 test periods are obtained from the test set in 2019 and sequentially denoted as 2019(S1), 2019(S2), 2019(S3) and 2019(S4).
The details of our datasets are listed in Table \ref{tab:data_des}. In summary, our training/validation set has more than a half million samples and is almost balanced. In addition, we have 72,545 samples in total for the test set, and each test period has more than 10,000 samples. We evaluate the prediction performance with the accuracy metric and employ binary cross entropy \citep{de2005tutorial} to measure the loss between $\hat{y}$ and $y$. 

\begin{table}[!ht]
\centering
\setlength{\belowcaptionskip}{10pt} 
\caption{\textbf{Summary statistics of CSI-300 for the training, validation and test datasets.} The dataset for training and validation has more than a half million samples and is almost balanced. The four seasons for test have 72,545 samples in total, and each season has more than 10,000 samples.}
\label{tab:data_des}
\resizebox{0.85\textwidth}{!}{%
\begin{tabular}{cccccccc}
\hline
\multicolumn{2}{c}{\multirow{2}{*}{Dataset}} & \multirow{2}{*}{\#Sample} & \multicolumn{2}{c}{$\downarrow$} &  & \multicolumn{2}{c}{$\uparrow$} \\ \cline{4-5} \cline{7-8} 
\multicolumn{2}{c}{}  &        & \#Sample & (\%)  &  & \#Sample & (\%)  \\ \hline
Train/Val                 & 2010\,-\,2018 & 530,284 & 276,533   & 52.15 &  & 253,751   & 47.85 \\ \hline
\multirow{4}{*}{Test} & 2019(S1) & 16,992  & 7,591     & 44.67 &  & 9,401     & 55.33 \\
                      & 2019(S2) & 17,765  & 9,534     & 55.36 &  & 8,231     & 46.33 \\
                      & 2019(S3) & 19,488  & 10,789    & 55.36 &  & 8,699     & 44.64 \\
                      & 2019(S4) & 18,300  & 9,142     & 49.96 &  & 9,158     & 50.04 \\ \hline
\end{tabular}}
\end{table}

In this paper, we select the optimal hyperparameters with grid search to obtain the best performance for all models. Specifically, we tune the sizes of the hidden representations within \{32, 64, 128, 256\} and the sizes of the minibatches within \{32, 128, 256\}. We use the Adam optimizer with an initial learning rate of $1e$-$3$. Following the computation flow in Algorithm \ref{code:prediction_module} and Figure \ref{fig:prediction_data_flow}, we employ the most popular deep learning library in research area, PyTorch which can automatically optimize learnable parameters through backpropagation, to implement our prediction module \footnote{The code of our framework is available at https://github.com/BUAA-WJR/PriceGraph}. For all baselines, we train the models in an end-to-end manner from raw quote data with a $z$-$score$ normalization function using the standard deviation and mean calculated for each sample. For the parameters in struc2vec, we set walk-length as 10 and window-size as 3 which are consistent with parameters set in \citep{ribeiro2017struc2vec}. The classical time series approaches, i.e., ARMA and GARCH, are optimized through $Auto$-$Arima$ in Python.

\section{Results}
\label{sec:results}
In this section, we show the empirical results obtained from numerical experiments. Further back-test experiments are also conducted to demonstrate the effectiveness of our proposed framework.

\begin{figure}[!ht]
  \centering
  \includegraphics[width=1.\textwidth]{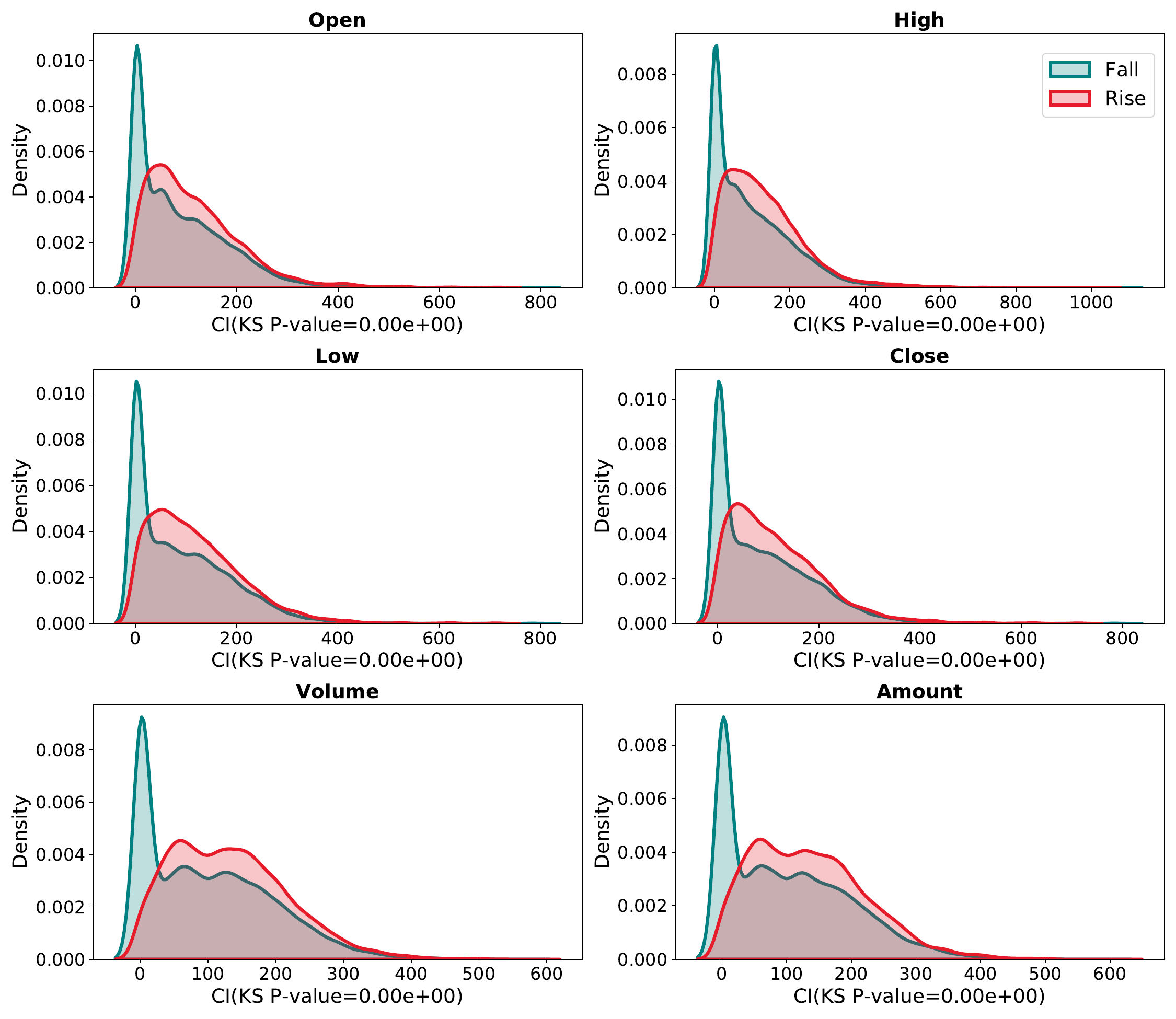} 
  \caption{\textbf{CI distributions of price graphs converted from a sample stock with six market price data from 2010 to 2019.} We also examine the two groups of distributions in the $Rise$ and $Fall$ directions by a Kolmogorov--Smirnov test (KS test), where the KS test p-values show that the two groups of distributions are distinct (i.e., all p-values are less than 0.001).}
  \label{fig:ci_dis}
\end{figure}

\subsection{CI distribution of price graphs}
To examine the quality of the price graphs obtained from the daily quote data of stocks, we take the CI distributions to assess whether the converted price graphs can capture the intrinsic properties of stocks and whether CIs are capable of distinguishing between graph nodes. Specifically, for each market price dataset, we split the converted price graphs into two groups by the corresponding target value and calculate the CI for each graph from 2010 to 2019. For graphs with target 1, we denote this group as $Rise$ and color them red. For graphs with target 0, we denote this group with $Fall$ and mark them green.
Figure \ref{fig:ci_dis} shows the CI distributions of a stock sampled from CSI-300 component stocks with the ticker symbol 002624. We find that the CI distributions of the two groups across the six market price datasets can be distinguished from each other. Furthermore, in addition to the visual differences, we also employ the Kolmogorov--Smirnov test to examine whether the two groups of distributions are significantly distinct in terms of statistics. We can see in Figure \ref{fig:ci_dis} that all $p$-values are less than 0.001; that is, the two groups of CIs have different distributions, which indicates that the price graphs converted from stock quote data can leverage certain intrinsic properties of stocks and separate rises from falls. Besides the CI distributions within two groups, we also inspect the discriminative ability of other structural indicators of nodes, such as degrees. However, they are not as distinguishable as CI within two groups, which indicates that CI can not only achieve more accurate or more discriminative attention weighing in model learning but also help model control the possible bias when facing abnormal nodes caused by chaotic property of financial time series. At the same time, this is consistent with the excellent performance of CI in influence description.

\begin{figure}[!ht]
  \centering
  \includegraphics[width=1.\textwidth]{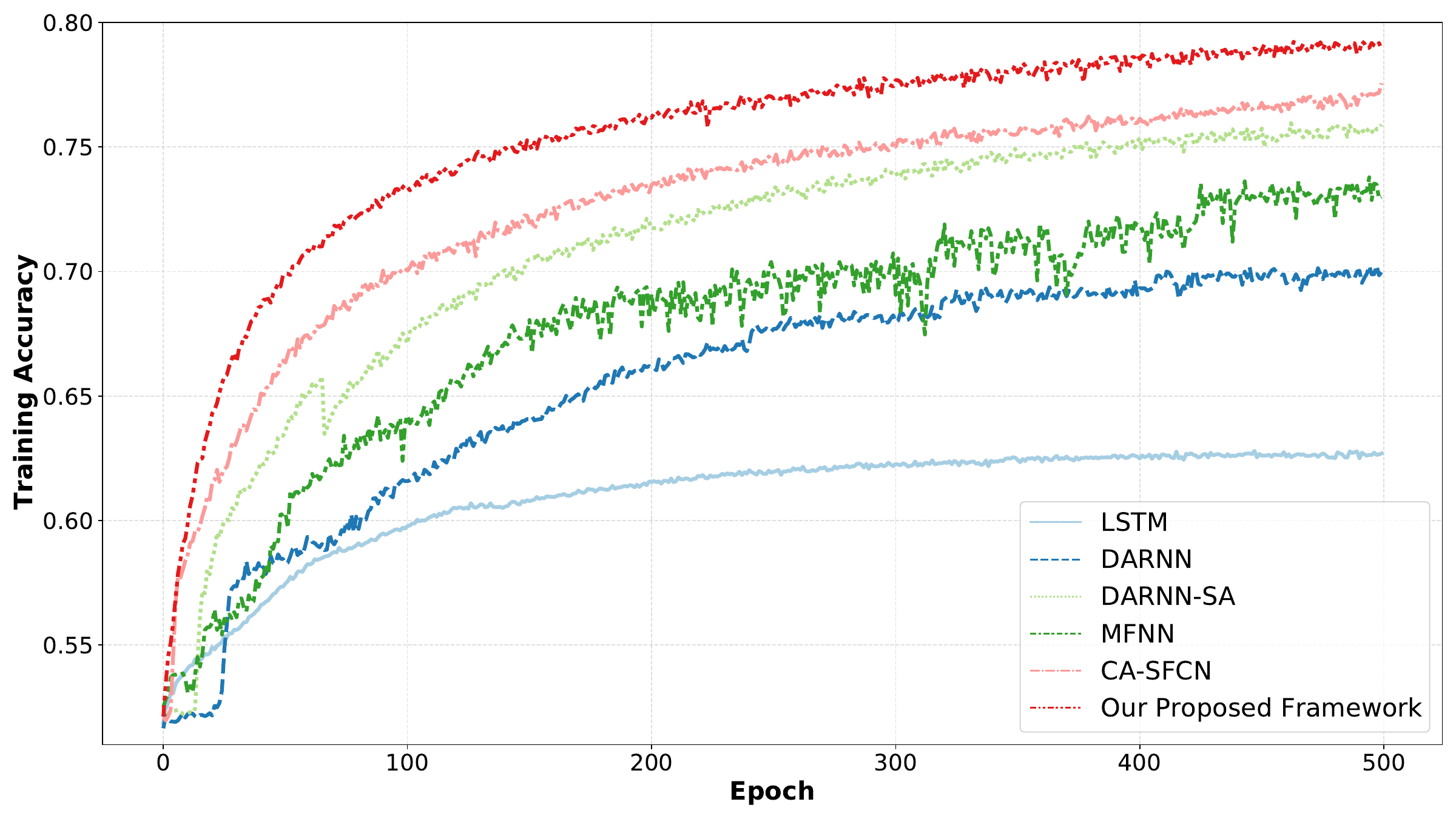} 
  \caption{\textbf{Training set performances of our proposed framework and baselines.}} 
  \label{fig:train_accs} 
\end{figure}

\subsection{Training set performance}
We further assess the representational power of our proposed framework by comparing its training accuracies with that of the baselines. Models with higher representational power should have higher training set accuracy. 
Figure \ref{fig:train_accs} shows the training accuracy curves of our proposed framework and all five baselines on the same training set from 2010 to 2018 \footnote{Although we train all models with longer epochs, Figure \ref{fig:train_accs} only shows the accuracies for the first 500 epochs because all models converge by this point.}.
First, our proposed framework is capable of obtaining the highest training accuracy, which implies that certain properties of stock quote data can be captured by the structural information derived from price graphs.
In comparison, despite utilizing the same learning component, DARNN-SA still cannot defeat our proposed framework. DARNN-SA yields a significant accuracy gain over the DARNN when fitting training data, which confirms the positive effect of the interrelationships between stocks \citep{wang2019alphastock}.
Compared with our proposed framework, CA-SFCN achieves competitive performance on the training dataset with only six market price data, which confirms the superior capability of component-based CNNs for capturing short-range dependencies \citep{hao2020new}.
In particular, LSTM severely underfits the training dataset.
In our experiments, in addition to feature engineering approaches, such as the CNNs in the MFNN and CA-SFCN, attention mechanisms also have remarkable predictive capabilities.

\subsection{Test set performance}

\begin{table}[!hb]
\centering
\setlength{\belowcaptionskip}{7pt} 
\caption{\textbf{Results (\%) of our proposed framework and the baselines.} All models predict price trend labels at the next time step. The best-performing results are highlighted with boldface. Our proposed framework outperforms all the state-of-the-art baselines on the test accuracies.}
\label{tab:test_metrics}
\resizebox{\textwidth}{!}{%
\begin{threeparttable}
\begin{tabular}{lcccccccccccccccc}
\hline
 & \multicolumn{4}{c}{2019(S1)} & \multicolumn{4}{c}{2019(S2)} & \multicolumn{4}{c}{2019(S3)} & \multicolumn{4}{c}{2019(S4)} \\ \cline{2-17}
 & Accuracy & Precision & Recall & F1 & Accuracy & Precision & Recall & F1 & Accuracy & Precision & Recall & F1 & Accuracy & Precision & Recall & F1 \\ \hline
ARMA & 50.15 & 54.96 & 42.81 & \multicolumn{1}{c|}{48.13} & 50.75 & 46.61 & 42.51 & \multicolumn{1}{c|}{44.46} & 49.89 & 44.26 & 39.91 & \multicolumn{1}{c|}{41.97} & 50.07 & 49.40 & 41.77 & 45.26 \\
GARCH & 50.28 & 54.90 & 44.68 & \multicolumn{1}{c|}{49.26} & 50.66 & 46.74 & 45.74 & \multicolumn{1}{c|}{46.23} & 50.43 & 45.05 & 41.76 & \multicolumn{1}{c|}{43.34} & 50.36 & 49.75 & 42.24 & 45.69 \\
LSTM & 57.94 & 64.88 & 52.26 & \multicolumn{1}{c|}{57.89} & 59.88 & 58.98 & 44.04 & \multicolumn{1}{c|}{50.43} & 56.71 & 52.23 & 35.33 & \multicolumn{1}{c|}{42.16} & 54.75 & 55.97 & 44.92 & 49.84 \\
DARNN & 60.87 & 68.27 & 54.69 & \multicolumn{1}{c|}{60.73} & 62.03 & 61.48 & 48.29 & \multicolumn{1}{c|}{54.09} & 60.62 & 58.10 & 61.33 & \multicolumn{1}{c|}{59.67} & 61.54 & 68.34 & 63.31 & 65.73 \\
DARNN-SA & 64.32 & 71.72 & 58.63 & \multicolumn{1}{c|}{64.52} & 66.23 & 66.60 & 54.40 & \multicolumn{1}{c|}{59.89} & 65.47 & 65.00 & 59.09 & \multicolumn{1}{c|}{61.9} & 65.63 & 72.34 & 68.92 & 70.59 \\
MFNN & 61.21 & 68.28 & 60.81 & \multicolumn{1}{c|}{64.33} & 63.00 & 65.30 & 51.40 & \multicolumn{1}{c|}{57.52} & 62.74 & 67.68 & 58.75 & \multicolumn{1}{c|}{62.9} & 64.69 & 67.19 & 57.52 & 61.98 \\
CA-SFCN & 65.51 & 72.82 & 60.10 & \multicolumn{1}{c|}{65.85} & 67.21 & 67.61 & \textbf{73.52} & \multicolumn{1}{c|}{70.44} & 66.10 & \textbf{77.81} & 68.32 & \multicolumn{1}{c|}{\textbf{72.76}} & 67.30 & 70.24 & 73.38 & 71.78 \\ \hline
\textbf{Our framework} & \textbf{67.48} & \textbf{75.24} & \textbf{61.45} & \multicolumn{1}{c|}{\textbf{67.65}} & \textbf{68.46} & \textbf{69.81} & 71.67 & \multicolumn{1}{c|}{\textbf{70.73}} & \textbf{68.34} & 67.86 & \textbf{73.77} & \multicolumn{1}{c|}{68.09} & \textbf{67.91} & \textbf{77.51} & \textbf{73.78} & \textbf{75.60} \\ \hline
\end{tabular}
$Notes.$ Precision, recall and the F1 measure are metrics calculated in the upward direction.
\end{threeparttable}}
\end{table}

Next, we compare the achieved test accuracies to further evaluate our proposed framework. Although the training results do not directly reveal the generalization capability of structural information, it is not a delusion to expect that our proposed framework with strong representational power can accurately capture certain properties and thus generalize well. Table \ref{tab:test_metrics} compares the test accuracies of our proposed framework and the state-of-the-art baselines. It can be concluded that the performance on the test set is in line with that on the training set; that is, our proposed framework consistently outperforms the state-of-the-art baselines on the test set, obtaining the best accuracy. This reveals that the structural information contained in stock time series is capable of addressing fundamental questions regarding long-term dependencies and the chaotic property. Compared with that of the second-best method, CA-SFCN, the recall rate of our proposed framework is lower in the second season of 2019, and the precision is lower in the third season of 2019. As for the classical time series approaches, ARMA and GARCH do not even achieve competitive performance with the base deep learning method (i.e., LSTM). Although high precision and recall are preferred in the Chinese stock market, which only allows longing on stocks, consistently high accuracies are still able to ensure that profitable predictions are generated by our framework. Furthermore, it is worth mentioning that our proposed framework yields significant improvements in all evaluation metrics over those of DARNN-SA with six market price data as inputs; this again emphasizes the effectiveness of the structural information obtained from price graphs and the proposed framework for stock prediction.

\section{Discussion}
\label{sec:discussion}
To further assess the performance of the proposed framework, the predictive powers of structural information and the node weights of price graphs are tested. Furthermore, a trading strategy is also implemented based on the signals produced by the prediction model to validate the profitability and stability of our framework in more realistic scenarios.

\subsection{Ablation analysis}
To test the effectiveness of structural information, the contribution of each type of market price data is inspected by reducing complexity while preserving the predictive capability of our framework.
Borrowing the idea of feature selection from traditional machine learning methods \citep{guyon2003introduction}, the training and computational time can be saved by reducing the size of input tensor for deep learning approaches.
The input of the proposed framework corresponds to the number of market price data types. Therefore, we conduct an ablation study with only one price graph embedding (PGE) derived from each corresponding time series of price data as input; these embeddings are denoted as Close-PGE, Open-PGE, High-PGE, Low-PGE, Amount-PGE and Volume-PGE.
To simply exhibit the effectiveness of structural information, accuracy, the most universal indicator in prediction tasks, is adopted to inspect the performance of each type of market price data \citep{wu2020predicting}.
The accuracies are shown in the upper panel of Table \ref{tab:ablation}. As can be seen, the performances of the models tested on of stock price embeddings, i.e., close, open, high and low prices, are consistently better than those of the models tested on amount and volume embeddings. This proves that price trend prediction can be better performed by models with more direct inputs of price.
In particular, the model of Close-PGE acquires the highest accuracies in this ablation analysis except for the second season of 2019.
These results also indicate that the structural information obtained from different types of market price data related to stock prices contains distinct predictive power for predicting price trends. The results from our proposed framework represent an increase in predictive power through the extraction of structural information from a variety of market price data types.

\begin{table}[!htbp]
\centering
\setlength{\belowcaptionskip}{7pt} 
\caption{\textbf{Ablation analysis accuracy of each learning model (\%).} The increases of accuracy after introducing node weights are presented in parentheses behind accuracies.}
\label{tab:ablation}
\resizebox{\textwidth}{!}{%
\begin{tabular}{lcccccccccccccccc}
\hline
 & \multicolumn{16}{c}{The effectiveness of structural information} \\ \cline{2-17} 
 & \multicolumn{4}{c}{2019(S1)} & \multicolumn{4}{c}{2019(S2)} & \multicolumn{4}{c}{2019(S3)} & \multicolumn{4}{c}{2019(S4)} \\ \cline{2-17} 
 & Accuracy & Precision & Recall & \multicolumn{1}{c|}{F1} & Accuracy & Precision & Recall & \multicolumn{1}{c|}{F1} & Accuracy & Precision & Recall & \multicolumn{1}{c|}{F1} & Accuracy & Precision & Recall & F1 \\ \hline
Close-PGE & 61.24 & 67.32 & 57.47 & \multicolumn{1}{c|}{62.01} & 63.69 & 64.57 & 66.31 & \multicolumn{1}{c|}{65.43} & 62.23 & 61.74 & 64.54 & \multicolumn{1}{c|}{63.11} & 60.54 & 66.82 & 67.15 & 66.98 \\
Open-PGE & 60.34 & 65.43 & 55.97 & \multicolumn{1}{c|}{60.33} & 63.17 & 63.73 & 59.42 & \multicolumn{1}{c|}{61.50} & 61.97 & 61.41 & 63.27 & \multicolumn{1}{c|}{62.33} & 59.33 & 64.33 & 63.54 & 63.93 \\
High-PGE & 60.89 & 66.47 & 56.41 & \multicolumn{1}{c|}{61.03} & 64.13 & 65.54 & 62.76 & \multicolumn{1}{c|}{64.12} & 60.36 & 59.58 & 55.76 & \multicolumn{1}{c|}{57.61} & 60.12 & 65.97 & 64.27 & 65.11 \\
Low-PGE & 61.59 & 66.12 & 57.57 & \multicolumn{1}{c|}{61.55} & 62.69 & 62.91 & 65.34 & \multicolumn{1}{c|}{64.10} & 60.83 & 60.09 & 64.17 & \multicolumn{1}{c|}{62.06} & 59.64 & 62.13 & 66.56 & 64.27 \\
Amount-PGE & 58.65 & 63.44 & 52.15 & \multicolumn{1}{c|}{57.24} & 60.27 & 60.22 & 63.96 & \multicolumn{1}{c|}{62.03} & 59.41 & 58.32 & 61.91 & \multicolumn{1}{c|}{60.06} & 58.47 & 63.14 & 64.85 & 63.98 \\
Volume-PGE & 59.15 & 63.79 & 52.86 & \multicolumn{1}{c|}{57.81} & 60.52 & 60.25 & 62.77 & \multicolumn{1}{c|}{61.48} & 58.37 & 59.13 & 57.35 & \multicolumn{1}{c|}{58.23} & 59.37 & 62.57 & 54.13 & 58.04 \\ \hline
 & \multicolumn{16}{c}{The effectiveness of node weights} \\ \cline{2-17} 
 & \multicolumn{4}{c}{2019(S1)} & \multicolumn{4}{c}{2019(S2)} & \multicolumn{4}{c}{2019(S3)} & \multicolumn{4}{c}{2019(S4)} \\ \cline{2-17} 
 & Accuracy & Precision & Recall & \multicolumn{1}{c|}{F1} & Accuracy & Precision & Recall & \multicolumn{1}{c|}{F1} & Accuracy & Precision & Recall & \multicolumn{1}{c|}{F1} & Accuracy & Precision & Recall & F1 \\ \hline
DARNN-NW & 61.33 (\textbf{+0.46}) & 68.76 & 57.33 & \multicolumn{1}{c|}{62.53} & 62.94 (\textbf{+0.91}) & 63.42 & 50.74 & \multicolumn{1}{c|}{56.38} & 61.89 (\textbf{+1.27}) & 58.11 & 63.86 & \multicolumn{1}{c|}{60.85} & 62.13 (\textbf{+0.59}) & 68.48 & 63.53 & 65.91 \\
DARNN-SA-NW & 65.67 (\textbf{+1.35}) & 73.43 & 57.49 & \multicolumn{1}{c|}{64.49} & 67.22 (\textbf{+0.99}) & 67.06 & 55.2 & \multicolumn{1}{c|}{60.55} & 65.64 (\textbf{+0.17}) & 64.47 & 62.36 & \multicolumn{1}{c|}{63.40} & 66.72 (\textbf{+1.09}) & 70.36 & 69.87 & 70.11 \\ \hline
\end{tabular}%
}
\end{table}

Next, we delve deeper into the effect of node weights obtained from price graphs in terms of mitigating the chaotic property of financial time series. Specifically, we conduct a battery of additional experiments based on the DARNN and DARNN-SA. As shown in Equation \ref{eq:decoder_attn_ci}, we update the temporal attention by adding node weights as knowledge-based attention over temporal points. Therefore, we add the same node weights to the second decoder stages of DARNN and DARNN-SA while retaining the original model input and six market price data. As shown in the lower panel of Table \ref{tab:ablation}, the two variants are denoted as DARNN-NW and DARNN-SA-NW, respectively. Compared to the original versions without node weights (see Table \ref{tab:test_metrics}), the adjusted models yield obvious increases in accuracies for the four seasons. Thus, we can conclude that by using node weights to enhance temporal weight learning, we can refine crucial information from chaotic series and obtain better or at least comparable prediction results.

\begin{figure}[!hb]
  \centering
  \subfigure[Stock prices]{ 
    \includegraphics[width=.46\textwidth]{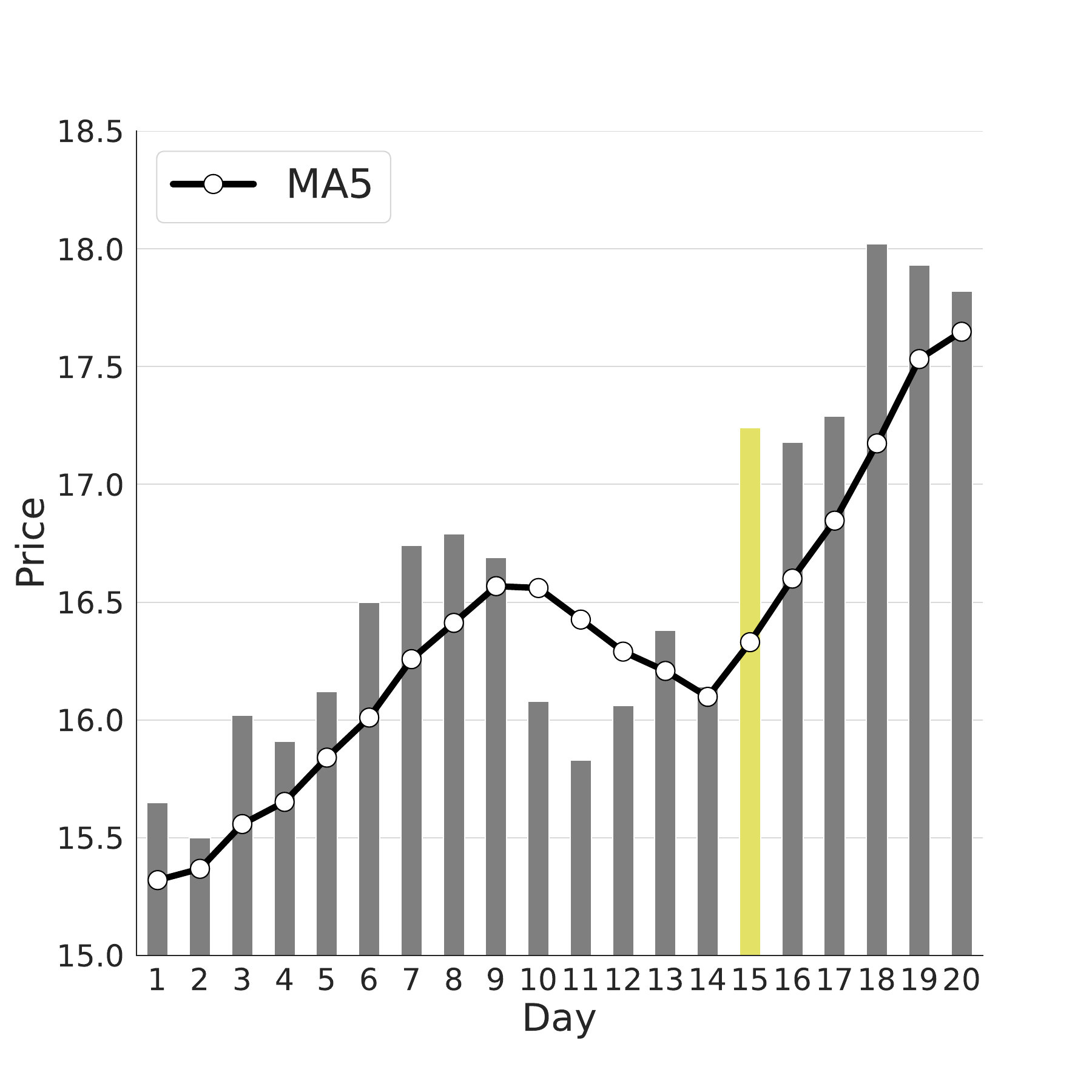} 
    \label{fig:ci_mec_price} 
  }
  \subfigure[Price graph with CI]{ 
    \includegraphics[width=.46\textwidth]{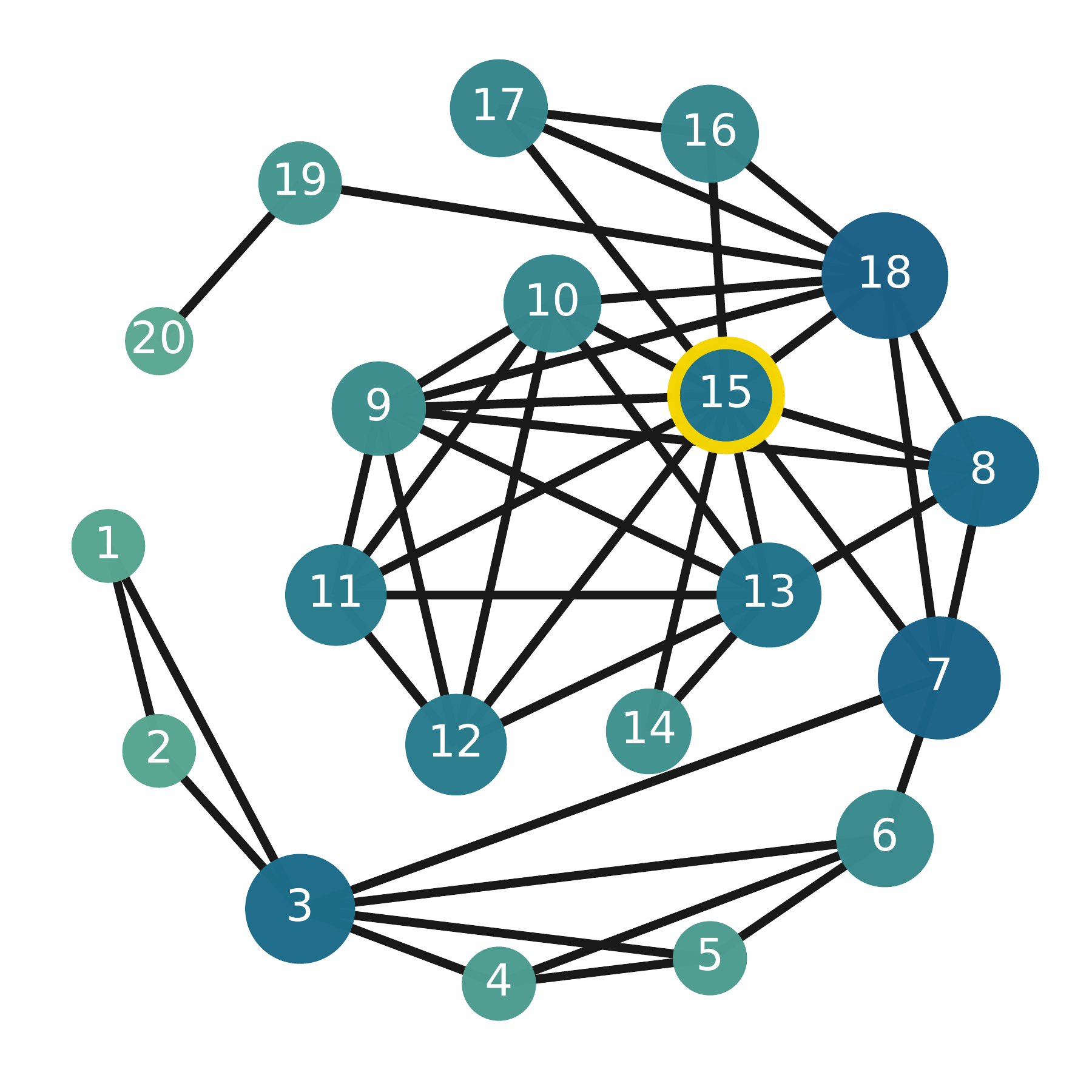} 
    \label{fig:ci_mec_vg} 
  } 
  \\
  \subfigure[Attention weights without CI]{ 
    \includegraphics[width=.46\textwidth]{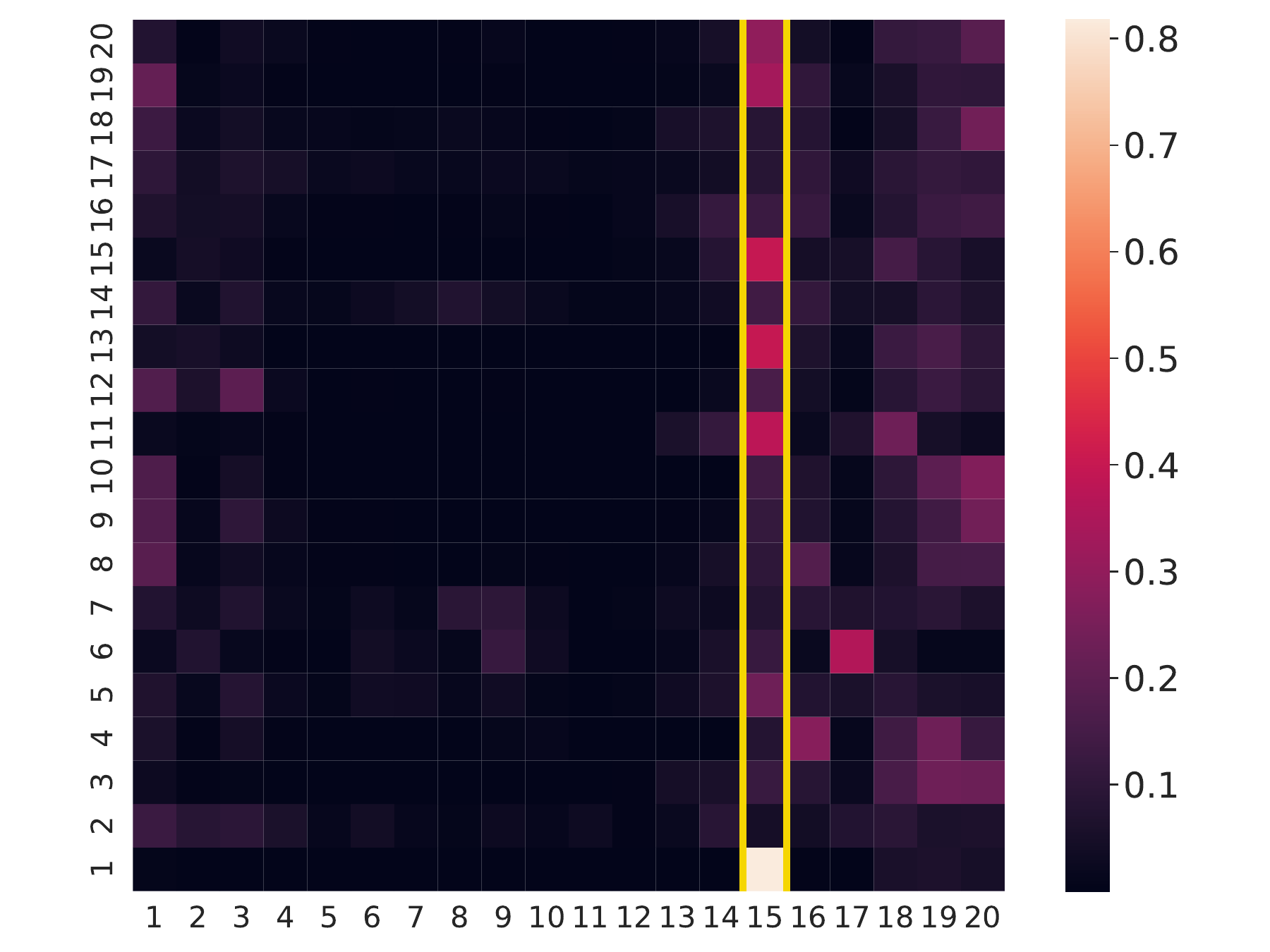} 
    \label{fig:ci_mec_withoutci} 
  } 
  \subfigure[Attention weights with CI]{ 
    \includegraphics[width=.46\textwidth]{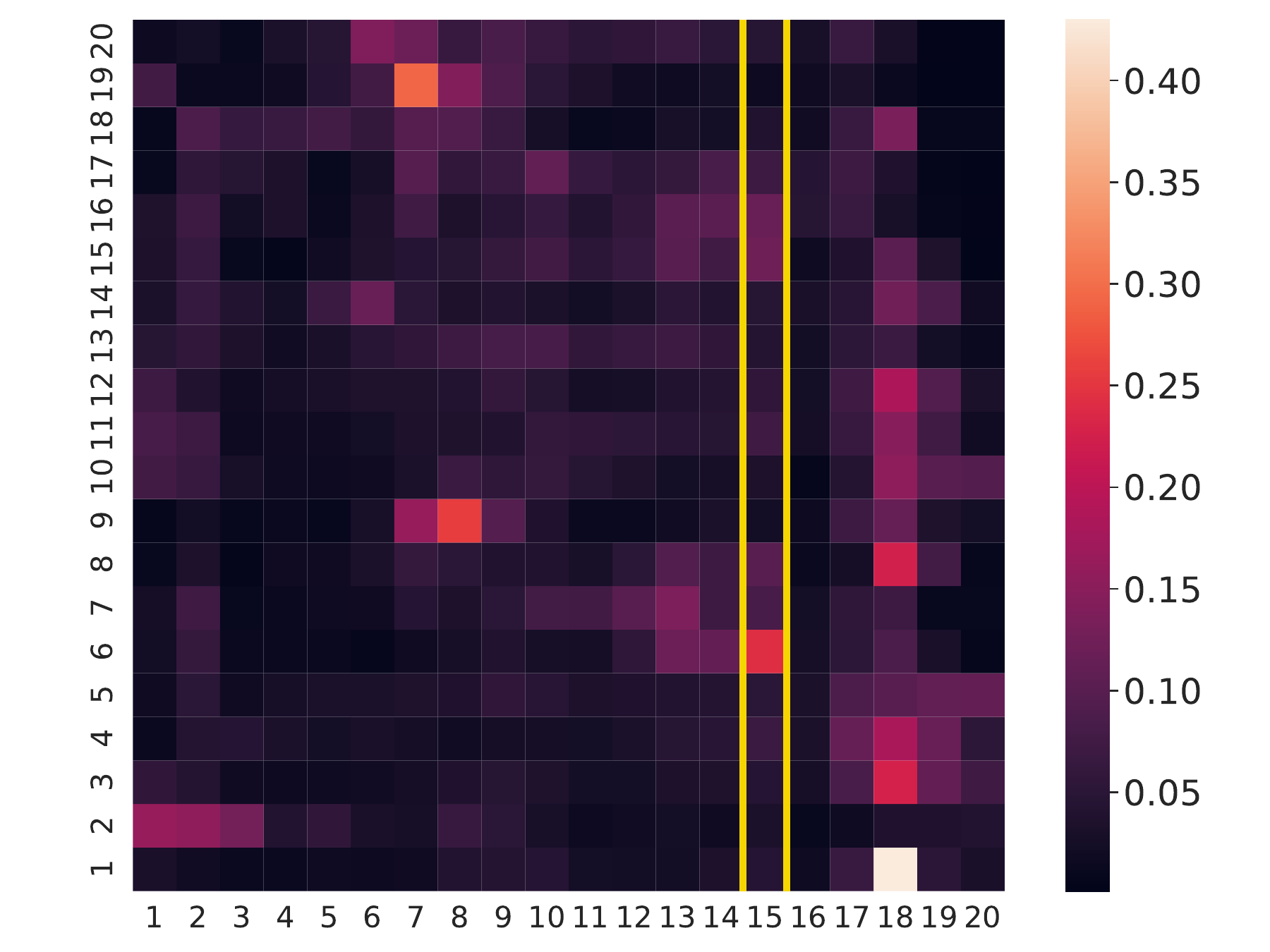} 
    \label{fig:ci_mec_withci} 
  } 
  \caption{\textbf{Illustration of how CI algorithm can help model capture real informative nodes.} Day 15 presents the largest price deviation from its five-day moving average (MA5), occupies the node with largest degree in corresponding price graph and attracts most of temporal attention of model without CI regularization (Equation \ref{temporal_tanh}). After updating the attention with CI as in Equation \ref{eq:decoder_attn_ci}, the learned model dispenses most temporal weights to Day 18 and also pays more attention to Day 8 and Day 7 instead of Day 15.}
  \label{fig:ci_mec}
\end{figure}

\subsection{Why CI works}
To obtain further insights into how CI overcomes the chaotic-property induced bias, a representative example of how the model decoder shifts its temporal attention from noisy points to real influential points is shown in Figure \ref{fig:ci_mec}. Given the chaotic property of financial series, stock prices are shown in non-stationary time series and have abrupt changes or unexpected reversals. Figure \ref{fig:ci_mec_price} shows such a typical price series, in which an abrupt price change or unexpected reversal appears at Day 15 and results in the largest price deviation from its five-day moving average. In addition, as can be seen in Figure \ref{fig:ci_mec_vg}, the Day 15 in the corresponding price graph also serves as the node of the largest degree. However, in Figure \ref{fig:ci_mec_vg}, the real informative (larger CI values) nodes are marked with deeper colors and greater sizes. Thus, we know that the Day 18 is actually the node with the greatest collective influence in this price graph followed by Day 8 and Day 7, suggesting indispensable signals they carry in the trending prediction of other nodes. In the process of temporal attention learning in Equation \ref{temporal_tanh}, we expect that the model could pay more attention to the real informative nodes that carry significant trending signals instead of abrupt changes. However, as can be seen in Figure \ref{fig:ci_mec_withoutci}, the model without CI regularization just follows the price change and is fully attracted by Day 15 rather than Day 18, implying the nodes of largest degrees resulted by abrupt changes in price would be wrongly targeted as informative ones in graph regardless of little trending information they actually carry. To fix this disadvantage of the vanilla model, we employ the CI of nodes to enhance the temporal weight learning process as in Equation \ref{eq:decoder_attn_ci}. As a result, Figure \ref{fig:ci_mec_withci} shows that the learned model agrees to dispense most temporal weights to Day 18 and also pays more attention to Day 8 and Day 7 rather than Day 15, which is indeed less informative for trending prediction.

\begin{figure}[!hb]
  \centering
  \includegraphics[width=1.\textwidth]{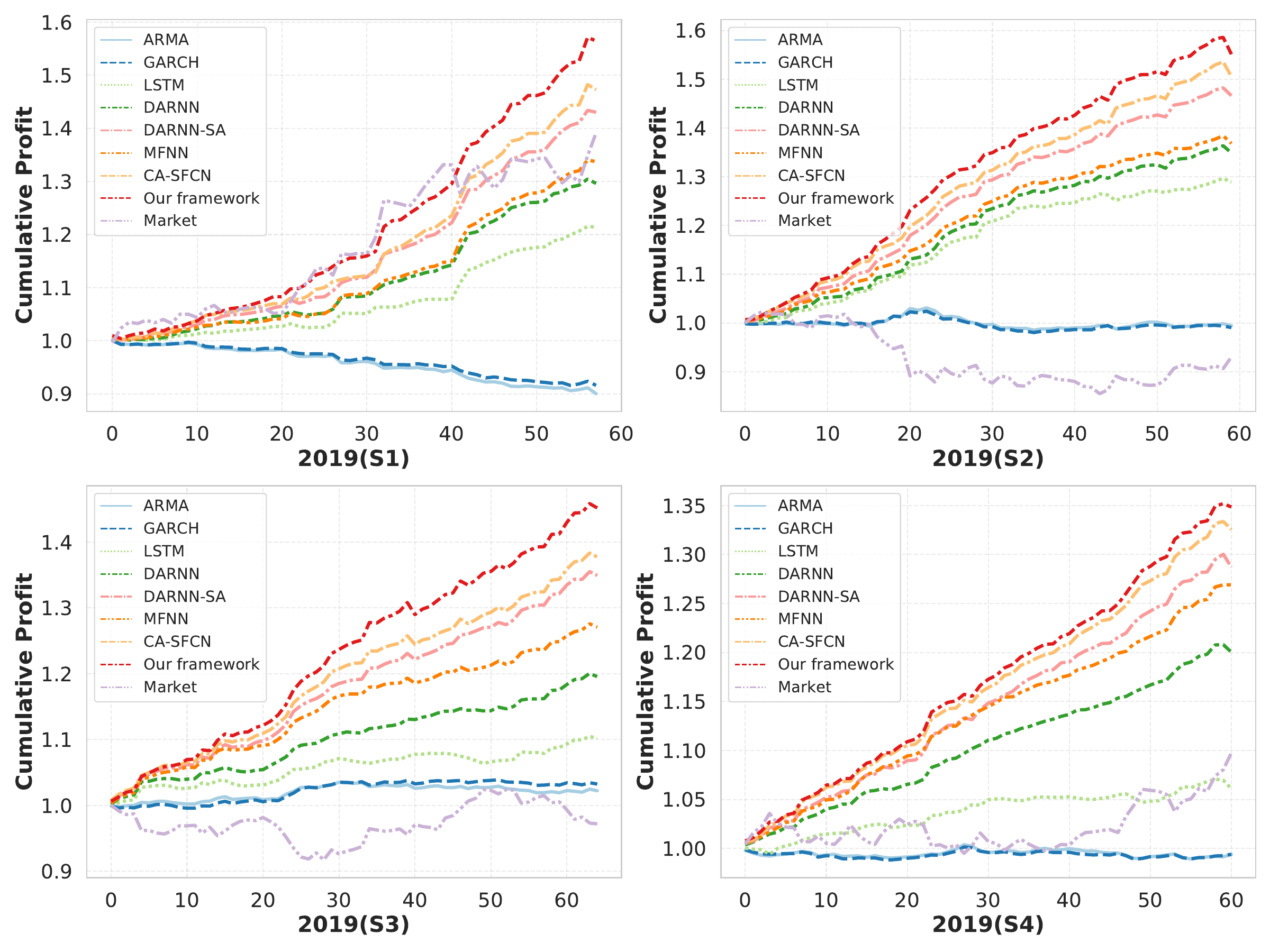} 
  \caption{\textbf{The cumulative profit curves of our proposed framework and different 
  baselines for the four seasons of 2019.}} 
  \label{fig:trading_profit} 
\end{figure}

\subsection{Trading simulation}
To further check the performance of our proposed framework, multiple back-tests are conducted by simulatively trading chosen stocks in our test set for the four seasons of 2019. The estimation strategy carries out trading with a daily frequency. In the process of simulation, according to the prediction of our framework, a simple rule is set to develop the trading strategy: if a rising trend of a stock price is given by our framework, we will take a long position that stock; while a falling trend of a stock price is predicted, we will take a short position for that stock. All stocks are evenly invested in and held for one day. In particular, all short/long operations open the position at closing price of the predicting day and close the position at closing price of the next day. Under the circumstance of no transaction cost, the cumulative profit are reinvested on the next trading day \citep{chen2019investment}. Although the transaction costs affect the final profit, the relative position based on model profit does not change owing to the use of the same trading strategy. We also calculate the average returns of the component stocks of CSI-300 by holding every stock evenly as the baseline, thereby indicating the overall market trend. All models' net value curves in the periods of simulations are shown in are shown in Figure \ref{fig:trading_profit}. Among all the baseline methods, our proposed framework gains the best profits (47.91\% return on average), even during the second season of 2019 when the market falls into a downturn. In particular, the market had a long rising period during the first season of 2019, and all of the models failed to gain enough profit to defeat the market, except for our framework, CA-SFCN and DARNN-SA.

In summary, all these results indicate that the structural information derived from price graphs by referring to the associations among temporal points and node weights can address the fundamental questions regarding long-term dependencies and the chaotic property of time series; moreover, the learned prediction model can help investors make more accurate trading decisions.

\section{Conclusion}
\label{sec:conclusion}
In this paper, based on time series graphs converted from market price data, we propose a novel framework to address fundamental questions by using structural information extracted from price graphs.
Through this framework, deep learning models collaborating with structural information achieve competent performance and display practical capabilities in stock prediction and trading. By employing models with attention mechanisms, we find that the long-term dependencies of the values in time series can be captured via structural information. Furthermore, by identifying prominent content from chaotic time series, models employing graph node weights as additional knowledge for temporal attention are capable of tackling the chaotic property of financial time series and achieving better stock prediction performance.
The results in this paper highlight the role of complex network methods for characterizing dynamical systems derived from time series. Compared with raw financial information such as market price data, structural information finely weaved from financial time series is verified to be superior for prediction tasks.

The superior performance of our proposed framework supplements the existing financial research on stock prediction by enriching the representations of time series through complex networks. Compared with previous studies that pay attention to networks among different stocks, sectors or markets \citep{xu2019interconnectedness, shahzad2021extreme, bouri2018does, ji2018network, shahzad2021impact}, the time series graphs based on temporal points from stock prices have more fine-grained information about the current states of stocks, which sufficiently helps us address the two fundamental questions regarding long-term dependencies and chaotic property in daily stock prediction. Notably, our framework is not limited to these observations mentioned in this study (i.e., market price data), other series related to price changes with time could also be new inputs to our framework. Accordingly, besides stock price prediction, our framework can also be extended and applied to other financial scenarios.

Our results also offer noteworthy implications for investors and policy makers. In this study, not only the transformation of financial series based on the VG algorithm but also the proposed deep learning model for financial series prediction have important inspirations for investors and supervisors in practice. Firstly, based on the structural information obtained from stock prices, our proposed framework outperforms the state-of-the-art models which take the raw market prices as input in testing accuracies. Additionally, the highest cumulative net values in trading simulation further highlight the effect of our framework in profit promotion. In this context, our work has important value in decision-making support for investors, especially institutional investors. Secondly, our work is also helpful for supervision. The superior performance of obtained results shows that the structural information of time series can be excellently preserved when performing complex network transformation, which makes it possible to depict the entire market structure, and provides a new angle for exploring propagation dynamics and early warnings of systemic risks.

Although the effectiveness of structural information for stock prediction is verified, there are also limitations in this study that inform the directions of future research. For example, in addition to graph embeddings, various characteristics of graphs that have been investigated in recent decades, such as graph centrality, clustering coefficients, and global efficiency, could also be informative for graphs and effective for stock prediction.
Moreover, feature engineering approaches, such as taking advantage of time series graphs, may lead to potential information loss during the process of structural information extraction. Therefore, given the ubiquity of graph neural networks in various time series problems, such as event forecasting \citep{deng2019learning}, transportation prediction \citep{wang2020evaluation} and recommendation \citep{huang2021multi}, an end-to-end learning model based on a graph neural network may be superior for financial time series prediction.
Both of the above limitations are promising directions for our future work.

\section*{Acknowledgments}
This work was supported by NSFC (Grant No. 71871006).


\end{document}